\documentclass[11pt]{article}
\usepackage[top=1.1in, bottom=1.1in, left=1.1in, right=1.1in]{geometry}
\AtBeginDocument{%
  \providecommand\BibTeX{{%
    \normalfont B\kern-0.5em{\scshape i\kern-0.25em b}\kern-0.8em\TeX}}}

\usepackage{amsmath}
\usepackage{algorithmicx}
\usepackage{array}
\usepackage{booktabs}   
\usepackage{subcaption} 

\usepackage{comment}
\usepackage{subcaption}
\usepackage{graphicx}
\usepackage{amsmath,amsfonts}
\usepackage{algorithm,algorithmicx,algpseudocode}
\usepackage{textcomp}
\usepackage{xcolor}
\usepackage{colortbl}
\usepackage{lipsum}
\usepackage{float}
\usepackage{pifont}
\usepackage{tabularx}
\usepackage{mwe}

\newtheorem{definition}{Definition}  

\begin{document}

\title{Learning Feature Interactions With and Without Specifications}

\author{Seyedehzahra Khoshmanesh \\ zkh@iastate.edu \and Tuba Yavuz \\ tuba@ece.ufl.edu \and Robyn R. Lutz \\ rlutz@iastate.edu}






\maketitle

\begin{abstract}
 Features in  product lines and highly configurable systems can interact in ways that are contrary to developers' intent. 
Current methods to identify such unanticipated feature interactions are costly and inadequate. To address this problem we propose a new approach to learn feature interactions, both in those product lines where constraints on feature combinations are specified and in  feature-rich configurable systems where such specifications often are not available.  The contribution of the paper is to use program analysis to extract feature-relevant learning models from the source code in order to detect unwanted feature interactions.
Where specifications of feature constraints are unavailable, our approach infers the constraints using feature-related data-flow dependency information.   Evaluation in experiments on three software product line benchmarks and a highly configurable system shows that this approach is fast and effective.  The contribution is to support developers by automatically detecting feature combinations in a new product or version that can interact in unwanted or unrecognized ways. This enables better understanding of latent interactions and identifies software components that should be tested together because their features interact in some configurations. 
\end{abstract}







\section{Introduction}\label{sec:introduction}

Feature-rich software systems are widely used when the goal is to satisfy a broad variety of customers.  Both software systems that form part of a product line and highly configurable software systems are feature-based, meaning that various combinations of features (units of functionality) are used to meet different needs \cite{apel3feature, berger2015feature}. The number of features and potential feature combinations tends to increase over time to accommodate the shifting needs of customers and users. \cite{pohl2005software,meinicke2016essential,mukelabai2018tackling}.

\begin{figure*}[th]
  \centering
  \includegraphics[width=\linewidth]{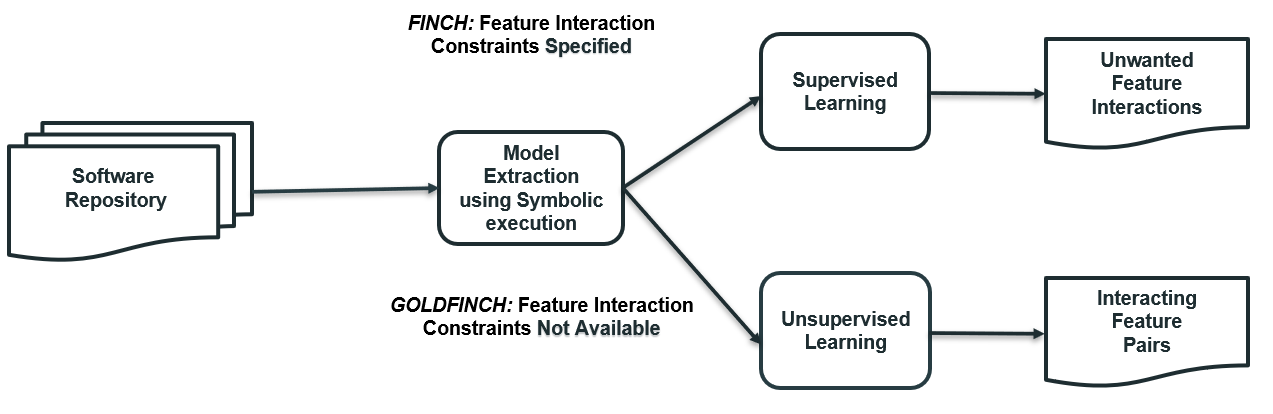}
  \caption{Overview of Our Proposed Approach to Learn Unwanted Feature Interactions whether or not Specifications of Feature-combination Constraints Are Available.}
  \label{fig:overview}
\end{figure*}

A {\it feature interaction} is an interference between two features such that, while each feature behaves as intended in a product with only one of the features present, having both features in a new product creates different, unintended behavior \cite{calder2003feature,apel2014feature}. 

An ongoing problem for developers of feature-rich systems is how to detect {\it unwanted feature interactions} in a new product or configuration \cite{apel2013feature}.  Unwanted feature interactions may involve new features or combinations of existing features that have not been used previously.  Unwanted feature interactions introduce unpredicted behavior, often escape testing, and add risk \cite{Lutz93, apel2011detection}. In safety-critical systems, they have caused multiple accidents
\cite{muscedere2019detecting,lutz2000software, apel2014feature,abdessalem2020automated}

Specification of constraints on feature combinations takes many forms, including feature models \cite{batory2020}, source code, user guides and operational manuals.  However, such documentation is often poorly maintained, partial, inconsistent with the code,  and quickly outdated \cite{nadi2014mining,nadi2015configuration, abal2018variability,soares2020feature}. Especially in highly configurable software systems, i.e., ones with features that can be added and removed \cite{northrop2002software, garvin2013failure}, specification of constraints on feature combinations often does not exist outside the code. 
We thus focus on specifications of feature constraints in the source code, as these are the specifications of greatest concern to developers and the most likely to be maintained and used.

The problem that this paper investigates is how to automatically identify unwanted feature interactions in a  software system whether or not specifications of constraints on feature combinations exist. 
To address this problem, \emph {we propose the use of symbolic-execution-guided machine learning to assist developers in identifying unwanted feature interactions} in a product line or feature-rich configurable system. Our approach is implemented and can automatically find problematic feature interactions and dependencies that may have been inadvertently introduced in a software version or new product. We envision that the most beneficial usage of our method will be to pinpoint for developers, prior to testing
, feature interactions of concern for  investigation and joint testing.

The need to identify unwanted or undetected feature interactions has driven multiple approaches to be proposed in the literature
 \cite{atlee2015measuring,abal2018variability,soares2018exploring,apel2021}.

Many approaches defer the effort until testing; however, despite important advances in sampling technologies, testing is inherently limited as a solution \cite{nguyen2016igen,temple2016using} 
and risks delaying discovery of  critical feature interactions until operations.   
Other approaches require the creation of formal models or other manually developed artifacts that most projects do not have \cite{atlee2015measuring}. Such problems are compounded in large configurable software systems where it is more likely that products are siloed, that each developer is familiar with only part of the system, and that documentation and configuration guidance are out of date

\cite{nadi2015configuration,Cashman18}. 
The limited adoption in practice of existing approaches motivates us to pursue automated learning of unwanted feature interactions from models extracted from source code.

Our goal in the work described in this paper is to provide automated support to learn unwanted feature interactions toward its use in developing feature-rich systems. 
Figure \ref{fig:overview} shows an overview of our approach. We use symbolic-execution-guided extraction of learning models from the code to identify unwanted feature interactions. 
As shown in the top half of the figure, our approach takes advantage of the fact that in software product lines, information from earlier products can be exploited to improve later products.  Here we describe a way to learn from knowledge of which features interfere with each other in order to identify, for a new product, problematic feature interactions that can occur in it. We call our approach and its implementation FINCH (Feature INteraCtion Help) as,  similar to miners' use of finch birds to detect poisonous carbon monoxide in coal mines, developers can use FINCH to detect and warn of unwanted or unknown feature interactions.  

However, many software product lines and most highly configurable systems lack constraint specifications or have only partial and/or obsolete specifications of feature constraints. The bottom half of Figure \ref{fig:overview} shows how we  generalize  our approach to handle  feature-rich systems without specifications of feature constraints. We call this extension of our approach and its implementation GOLDFINCH (Generalized FINCH).  All artifacts, code, and analysis used in this study are available at https://tinyurl.com/ydbtsc8r.

We evaluated the effectiveness of our approach by using it to learn feature dependencies in three product lines and a highly configurable subsystem.  Our evaluation addressed the following research questions:  
\begin{itemize}
\item \emph {RQ1:  How effective is the proposed approach in producing accurate results for known feature interactions?}

\item \emph{RQ2:  How accurately can the proposed approach predict new unwanted feature interactions based on existing feature interactions?}  

\item \emph{RQ3:  How scalable is the performance of our approach when specifications are unavailable?}

\end{itemize}

The contributions of the paper are:
\begin{itemize}
\item Where constraints on allowable feature combinations are specified, we learn unwanted feature interactions, first  using symbolic execution to extract a model from prior products' normal and failed paths and then using it to classify unseen paths in a new product.   
We show that this approach is successful in identifying even some unseen feature interactions and even with partial data.

We generalize our approach to highly configurable systems, where constraints on allowable feature combinations typically are not specified.  We use dynamic symbolic execution to detect feature-related data flow dependencies and association rule mining to infer feature interactions.  
 
We show that our method locates feature interactions of concern to developers.
\item We present evaluation results
showing that our method is fast and effective in detecting unwanted feature interactions.  With 
specifications, FINCH predicted all 19 known unwanted feature interactions in three product lines within seconds.  Without access to specifications, our generalized GOLDFINCH method 
still could detect 75-100\% of these unwanted feature interactions 
and, in a configurable system with 139 features, identified a data flow dependency between two features that previously were involved in an unwanted feature interaction.

\item We describe how our method supports developers by automatically finding features that interact in unrecognized or unwanted ways, and thus should be analyzed and tested together. 
\end{itemize}

\begin{figure}[th!]
\centering
\includegraphics[width=3.5in]{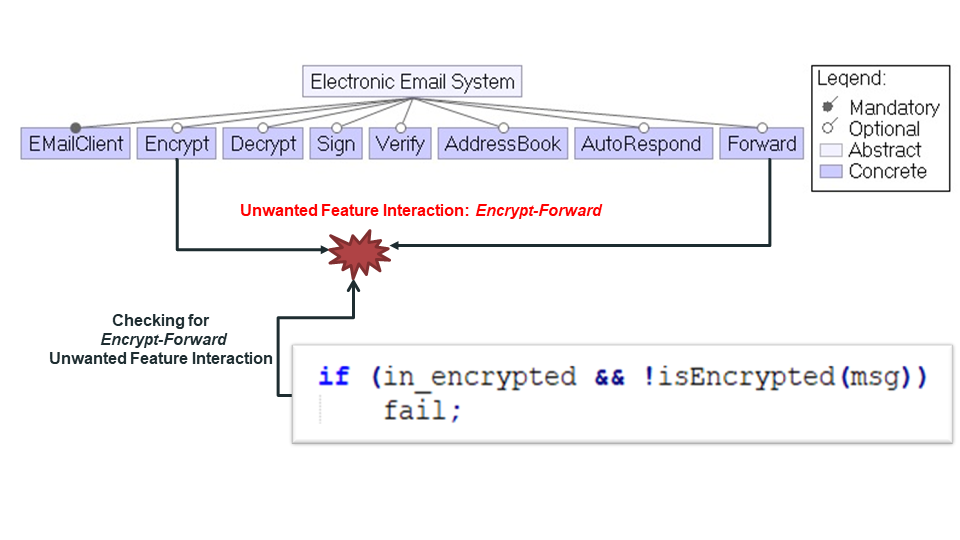}
\caption{Email Product Line Example with Unwanted Feature Interaction Specification.}
\label{fig:motivation_email}
\end{figure}

The remainder of the paper is organized as follows. Section \ref{sec:motiv} introduces two motivating examples.   Section \ref{sec:modex} describes our feature-relevant model extraction approach. Section \ref{sec:learning} describes how we learn feature interactions and introduces the case studies used in the experiments. Section \ref{sec:results} describes results from experiments evaluating our approach in light of our research questions. Section \ref{sec:related} describes related work. Section \ref{sec:conclusion}  discusses use cases and threats to validity, and Section \ref{sec:summary} offers a brief summary of contributions and concluding remarks.

\section{Motivating Examples}
\label{sec:motiv}

In this section, we introduce two examples of unwanted feature interactions that illustrate the problem and challenges that motivated our approach. 

\subsection{An unwanted feature interaction in a system with feature-constraint specifications}
The first example is from an electronic email system where a feature-constraint specification exists that prohibits two optional features from both being included in a product.  This is done by checking in the code whether the unwanted feature interaction occurs or not.

The Email software product line is a well-known benchmark in the literature \cite{hall2005fundamental,apel2011detection}.  As shown in Figure \ref{fig:motivation_email}, the Email product line has a base Email Client shared across all products as well as seven optional features.  
 There are ten known unwanted feature interactions in the Email case study from \cite{colder2000feature,apel2011detection}

A constraint specification in the code checks two conditions: if the second party in the scenario received the email as encrypted,  and if the message in the email is kept as encrypted when forwarding it to the third party.  A failure occurs when the email is received as encrypted, $in\_encrypted$, but is not kept as encrypted when forwarding to the third party, $!isEncrypted(msg)$. This happens because the second party does not have the third party's key and therefore forwards the email's message as plain text to the third party, violating the system's security property. 
Our proposed approach to automating unwanted feature interaction detection uses feature-constraint specifications, when they are available,
to learn the failure paths and normal paths in the various products of the product line.  

\subsection{An unwanted feature interaction in a system without feature-constraint specifications}

\begin{figure}[th!]
\centering
\includegraphics[width=9cm]{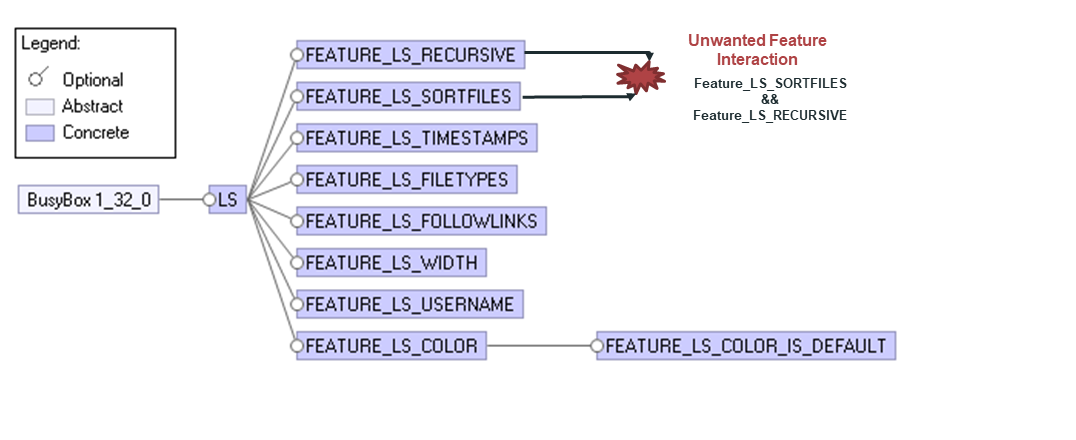}
\caption{The Vulnerability in the coreutils/ls Example.}
\label{fig:motivation_busybox}
\end{figure}

\begin{figure}[th!]
\begin{center}
\begin{footnotesize}
\begin{verbatim}
%\begin{minted}[mathescape,
%               linenos,numbersep=2pt,autogobble,xleftmargin=0.03\textwidth]
%               {c}
#ifdef CONFIG_FEATURE_LS_RECURSIVE
void dfree(struct dnode **arr)
{
  cur = arr[0]; // is no longer the head
  while (cur != NULL) {
    next = cur->next;
    free(cur); // Load
    cur = next;
  }
}
#endif

void showdirs(int **arr)
{
 ...
#ifdef CONFIG_FEATURE_LS_RECURSIVE
  dfree(arr); //ERROR
#endif
}

#ifdef CONFIG_FEATURE_LS_SORTFILES
void sort(int **arr, int size)
{
...
  for(i=0;i<size;i++) {
    for(j=i;j<size;j++) {
      if(*arr[i] > *arr[j]) {
        temp=*arr[i];
        *arr[i]=*arr[j]; // Store
        *arr[j]=temp;
      }
    }
  }
}
#endif

int main(int argc, char **argv)
{
#ifdef CONFIG_FEATURE_LS_SORTFILES
    sort(arr, size);
#endif
    showdirs(arr);
}               
%\end{minted}
\end{verbatim}
\end{footnotesize}
\end{center}
\caption{A Code Excerpt Demonstrating a Memory Leak Bug Related to Interaction of the Features {\tt LS\_SORTFILES} and {\tt LS\_RECURSIVE}.}
\label{fig:motivls}
\end{figure}            

For many real-world software product lines and most configurable systems, disallowed combinations of features or configuration options are not specified or are not available to developers \cite{soares2018exploring}. Our second example, which motivates us to investigate unwanted feature interactions in those cases {\it without} feature-constraint specifications, is an unwanted feature interaction in BusyBox, a large, highly configurable system.  BusyBox is a collection of many Unix utilities implemented in a single executable file \cite{wells2000busybox}. It has over 1000 configuration options or features \cite{abal2018variability}. 

Figure \ref{fig:motivation_busybox} shows the features in 
the ls part of coreutils in BusyBox 1.32.0. The ls utility is used to list files or directories in Linux and other Unix-based operating systems. As shown in Figure \ref{fig:motivation_busybox}, there are nine optional features in ls that a user can enable. 
However, an unwanted feature interaction occurs if both features {\tt LS\_SORTFILES} and {\tt LS\_RECURSIVE}  are enabled.

Figure \ref{fig:motivls} shows a code excerpt demonstrating how two features, {\tt LS\_SORTFILES} and {\tt LS\_RECURSIVE}, interact in a way that leads to a memory leak. The bug is manifested when 
both features are enabled. The root cause of the memory leak is a data-flow dependency between some program statements within the {\tt sort} and {\tt dfree} functions. 
The assumption that {\tt arr[0]} refers to 
the head of the link list gets invalidated by the {\tt sort} function.
An example data-flow dependency  
in this feature-controlled code 
is the store instruction at line 29 and the load instruction at line 7.
This bug is reported in the BusyBox bug database and classified as a variability bug in \cite{abal2018variability}. Variability bugs are bugs related to features or configuration options in highly configurable systems \cite{abal2018variability}. A variability bug with two or more features contributing to it is a feature interaction.

Since multiple studies have shown that most feature interactions involve two features, our aim is to detect pairwise unwanted feature interactions \cite{williams2000determination,oster2010automated, liebig2010analysis,siegmund2012predicting,siegmund2012spl,soares2018exploring,kolesnikov2019relation,abal2018variability}.
In our proposed approach to detecting unwanted feature interactions in the absence of constraint specifications, we use feature-related data flow dependencies to learn feature dependencies and present that information to developers.

\section{Feature Relevant Model Extraction}
\label{sec:modex}

In this section, we present extraction of various feature relevant models from the source code using program analysis. 
Section \ref{sec:featuremodels} presents the types of models we use,
and Section \ref{sec:extraction} explains their extraction using symbolic execution.

\subsection{Source Code Level Feature Models}
\label{sec:featuremodels}

Figure \ref{fig:overview} in Section \ref{sec:introduction} showed an overview of our approach. The first step toward predicting unwanted feature interactions, described here, is to extract the feature-relevant models we need from the code. 

\subsubsection{Control-Flow Models}
\label{sec:cfmodels}

The control-flow information on
an execution path can be described in various ways. 
One basic way is to describe it in terms of the sequence of executed 
instructions. Although this model would be very detailed, it is 
difficult to make associations with the features at the level 
of instructions without additional context information. 
Another way is to describe control-flow in terms of the set of stack traces that get generated. A {\it }stack trace provides a snapshot of a possible way of reaching a program location in terms of a sequence of callsites and provides context information at the level of functions and code locations. Features are often implemented by a set of
functions or they do interact in the context of certain functions. 
Therefore, a stack trace can potentially provide information 
relevant to a single feature or to the interaction of multiple features.
In this work, we use use stack trace information as a control-flow model.

\subsubsection{Data-flow Models}
\label{sec:dfmodels}

A data-flow model provides information about the values of program 
variables at various program locations and how these values flow from 
one program variable to another. 
Variables take different roles depending on whether they are input variables, other variables used internally to carry computations, or output variables.

Input variables may directly or indirectly relate to a feature. 
Therefore,  values of the input variables may control
 how a feature behaves individually as well as 
how features behave collectively. 

Additionally, the data-flow between two variables  
may determine 
observable variations in behavior that occur when certain features are 
enabled or disabled. We consider two types of dependencies: 
{\em store-load} and {\em store-store}.

\begin{definition}[Store-Load]
There is a store-load dependency between a store instruction $s$ and 
a load instruction $l$ if they both access the same memory region, $s$ happens before $l$, and $l$ reads the value stored by $s$, i.e., there 
is no other intervening store between $s$ and $l$.
\end{definition}

\begin{definition}[Store-Store]
There is a store-store dependency between two store instructions $s_1$ 
and $s_2$ if they both write to the same memory region without any 
other intervening write to that memory region between the first one and the second one.
\end{definition}

In this work, we thus extract and use two types of data-flow models: 1) data constraints  on the feature-related input variables, and 2) the store-load and store-store dependencies of the feature-related variables.

\subsection{Extracting Feature Models}
\label{sec:extraction}

We use \emph {dynamic symbolic execution}  to extract our feature models due to its ability to 
provide the stack-trace-based control-flow model and 
the three 
types of data-flow dependencies described in Section 
\ref{sec:featuremodels}.

We use the PROMPT tool \cite{yavuz2020analyzing}, which is designed to work at the component level. PROMPT extends the KLEE symbolic execution engine \cite{cadar2008klee} using lazy initialization and, therefore, can be applied at the function-level without the need for a 
test-driver. This enables extraction of feature relevant models at 
the component-level. 

In dynamic symbolic execution, inputs are labeled as symbolic to 
represent the fact that they can take any values. 
The underlying symbolic execution engine keeps the memory as a 
map from addresses to values that can be concrete values or 
symbolic expressions. 
The engine interprets each instruction according to 
its operational semantics. If all the operands have concrete values, then
the standard semantics is applied. 
If any of the operands has a symbolic expression as its value then the engine either produces another symbolic 
expression as a result (for non-branching instructions) or uses the symbolic expression to decide the feasibility of the branch targets (for branching instructions). Branching due to symbolic expressions may also happen in non-branching instructions such as load and store if the address is a symbolic expression. In that case, a separate path is generated for each memory object that may correspond to the symbolic 
expression. So, dynamic symbolic execution generates 
a symbolic execution tree, where the leaf nodes  correspond to 
different execution paths  
and the internal nodes with multiple children represent branching decisions due to symbolic expression evaluation.
Symbolic execution explores the underlying program for 
feasible paths.  
However, due to the well-known path explosion problem, symbolic execution may fail to explore all possible paths 
In this work, we apply symbolic execution at the component-level 
as implemented in the PROMPT tool \cite{yavuz2020analyzing} to deal with the path explosion problem.

Algorithm \ref{alg:baselinesymex} shows the baseline 
symbolic execution algorithm as implemented in KLEE. 
It keeps track of the set of active paths $active$, which gets initialized with the initial state of a given program $P$.
As long as there are some active paths and the timeout $\tau$ has 
not been reached, it executes 
the next instruction $inst$ on the current path $s$ and updates $active$ with the successors of $s$, which is denoted by a call to $\mathit{executeAndUpdate}$. 
Branch instructions 
and memory access instructions (load/store) with symbolic addresses 
may generate multiple successors and lead to the creation of new 
paths, which become the children of $s$ in the generated symbolic execution tree. 

Another important side-effect of the $\mathit{executeAndUpdate}$ operation is to update the path condition on the current path $s$ 
as well as on its successors. The path condition, $\mathit{PC}$ is the conjunction 
of all the symbolic constraints that have been found to be feasible on 
the current path. These constraints originate either from the conditions of the branching instructions or the constraints on the symbolic addresses, e.g., whether a symbolic index expression evaluates to a valid range or not for a given memory region, which is modeled as an array of bytes in KLEE. The feasibility of these constraints 
are checked using an SMT solver. 

For this project we have extended the symbolic execution engine: 1) to record the 
stack traces as a sequence of callsites and the path constraints as a set of atomic constraints, 2) to distinguish paths with respect to normal termination versus termination with an error, and 3) to compute store-load and store-store dependencies.

\begin{algorithm}
\caption{Baseline Symbolic Execution Algorithm.}
\label{alg:baselinesymex}
\begin{footnotesize}
\begin{algorithmic}[1]
\State {\bf BaselineSymEx}($P$: $PL$, $\tau$: $\mathcal{Z}$): 
\State $active \gets \{ \mathit{initState(P)} \} $
\While{$active \not = \emptyset$ AND $\tau$ not reached} 
   \State $s \gets \mathit{chooseNextState}(active)$
   \State $inst \gets nextInst(s)$
   \State $\mathit{executeAndUpdate}(s, inst, active)$
\EndWhile
\end{algorithmic}
\end{footnotesize}
\end{algorithm}

Algorithm \ref{alg:symex} shows how we extend the baseline symbolic execution algorithm to detect feature-relevant interactions. Specifically, we introduce metadata that stores control-flow and 
data-flow dependencies and introduce Algorithm \ref{fig:executeTrack} to extend the $\mathit{executeAndUpdate}$ algorithm that performs symbolic execution of an instruction within the context of a 
symbolic execution path. 
To keep track of the terminated paths according to their termination status, we detect instructions that cause termination or an error and 
update the relevant metadata, $\mathit{failTerm}$ or $\mathit{normalTerm}$, which denote the set of paths terminated with an error and those that terminated normally, respectively. 

As shown in Algorithm \ref{fig:executeTrack}, at every callsite, we compute the current call sequence and update the 
set of call sequences for the current path in the map, $T$.
$SM$ keeps track of the most recent store on a memory object for each state.
For each store instruction, we record the store-store dependency with the most recent instruction that wrote into the same memory region in $SS$, if any. The algorithm also updates the most recent store to the relevant memory region in $SM$. 
For each load instruction, it records the store-load dependency 
with the most recent instruction that stored into the same memory region  in $SL$, if any.

\begin{algorithm}
\caption{Symbolic Execution Based Extraction of Control-flow and Data-flow Dependencies.}
\label{alg:symex}
\begin{footnotesize}
\begin{algorithmic}[1]
\State {\bf ExtractFeatureModels}($P$: $PL$, $\tau$: $\mathcal{Z}$, $L$: $\mathcal{Z}$): 
\State $normalPaths, failPaths$: 
$\mathcal{P}((\mathcal{P}(Sequence),\mathcal{P}(Constraint))$ \Comment{Normal/failure termination path models}
\State $SM$: $State \times Address \mapsto Instruction$  \Comment{Store Map}
\State $SS$, $SL$: $\mathcal{P}(Instruction \times Instruction)$ \Comment{Store-store/Store-load pairs}
\State $T: State \times Z \times \mathcal{P}(Sequence)$
\State $SM \gets \lambda s.a. \text{undefined}$
\State $SS \gets SL \gets \emptyset$
\State $T = \lambda s.l. \emptyset$
\State $active \gets \{ \mathit{initState}(P) \} $
\State $normalTerm, failTerm \gets \emptyset$
\While{$active \not = \emptyset$ AND $\tau$ not reached} 
   \State $s \gets \mathit{chooseNextState}(active)$
   \State $inst \gets \mathit{nextInst}(s)$
   \State $\mathit{executeTrackAndUpdate}(s, inst, active, normalTerm, failTerm)$
\EndWhile
\State $normalPaths, failPaths\gets \emptyset$
\For{$s \in normalTerm$}
  \State $normalPaths \gets normalPaths \cup \{(chooseLongest(T(s), L),$ $Atomic(s.PC))\}$
\EndFor
\For{$s \in failTerm$}
   \State $failPaths \gets failPaths \cup \{(\{callSeq(s.stack)\}, Atomic(s.PC))\}$
\EndFor
\State $SSPairs \gets \bigcup_{s \in normalTerm \cup failTerm} SS(s)$
\State $SLPairs \gets \bigcup_{s \in normalTerm \cup failTerm} SL(s)$
\State {\bf return} $(normalPaths, failPaths, SSPairs, SLPairs)$
\end{algorithmic}
\end{footnotesize}
\end{algorithm}

\begin{algorithm}
\caption{The Extended Instruction Execution for Tracking Data-flow.}
\label{fig:executeTrack}
\begin{footnotesize}
\begin{algorithmic}[1]
\State {\bf executeTrackAndUpdate}($s$, $inst$, $active$, $normalTerm$, $failTerm$)
\State $\mathit{executeAndUpdate}(s, inst, active)$
\For{each successor $s'$ of $s$ do} 
   \If{$inst$ is a callsite for function $F$} 
        \State $cseq \gets add(callSeq(s'.stack),F)$
        \State $l \gets length(cseq)$ 
        \State $T = T[s' \rightarrow T(s')[l \rightarrow T(s')(l) \cup \{cseq\}]]$
  \EndIf
  \If{$s'$ terminates with an error}
     \State $failTerm \gets failTerm \cup \{s'\}$
   \EndIf     
  \If{$s'$ terminates normally}
     \State $normalTerm \gets normalTerm \cup \{s'\}$
  \EndIf
  \If{$inst \equiv store \ V \ to \ A$}
     \State $SS(s') \gets SS(s') \cup \{(SM(s',m.baseAddress), inst)\}$
     \State Let $m$ denote the memory object that the store address $A$ corresponds to on path $s'$
     \State $SM(s',m.baseAddress) \gets inst$
  \EndIf     
  \If{$inst \equiv load \ A$}
     \State $SL(s') \gets SL(s') \cup \{(SM(s',m.baseAddress), inst)\}$
     \State Let $m$ denote the memory object that the store address $A$ corresponds to on path $s'$      
  \EndIf
\EndFor  
\end{algorithmic}
\end{footnotesize}
\end{algorithm}

\subsection{Examples of Extracted Models}
\label{sec:modelexamples}

In this section, we provide examples of the models that we extracted 
from three software product-line benchmarks and the BusyBox configurable software system. 

\begin{figure}[!h]
  \centering
  \includegraphics[scale=0.6]{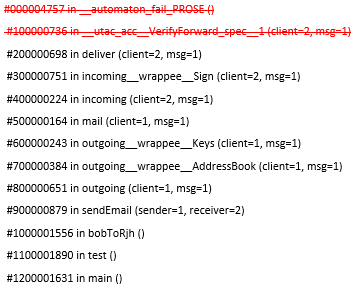}
  \caption{Stack Trace of a Verify-Forward Feature-Interaction Failure Path}
  \label{fig_approach_stack_failure}
\end{figure}

\begin{figure}[!h]
  \centering
  \includegraphics[scale=0.4]{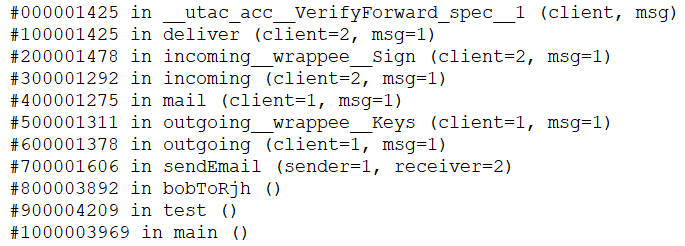}
  \caption{Stack Trace of a Normal Path of Verify-Forward }
  \label{fig_approach_stack_normal}
\end{figure}

Figures \ref{fig_approach_stack_failure} and \ref{fig_approach_stack_normal} show examples of control-flow models in the form of stack traces for a failure path and a normally terminated path, respectively. The red lines in Figure \ref{fig_approach_stack_failure} 
indicate some post-processing we perform to clean the data. In order to remove bias, we exclude function calls, e.g., {\tt \_\_automaton\_fail} to check the feature-constraint specifications in the code. 

\begin{figure}[!h]
  \centering
  \includegraphics[scale=0.4]{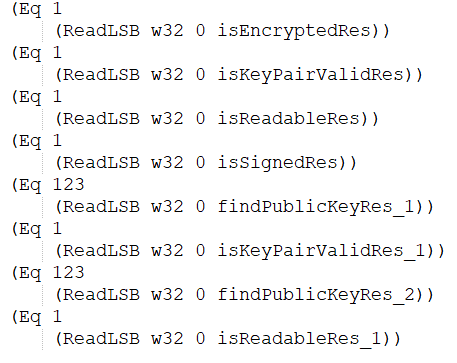}
  \caption{Path Constraints of a Verify-Forward Unwanted Feature-Interaction Path in KQuery Format.}
  \label{fig_approach_path_failure}
\end{figure}

\begin{figure}[!h]
  \centering
  \includegraphics[scale=0.4]{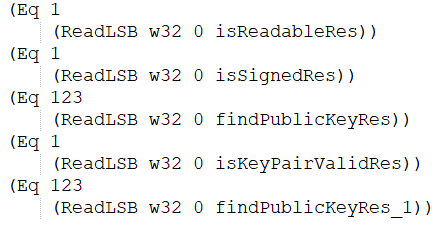}
  \caption{Path Constraints of a Normal Path of Verify-Forward }
  \label{fig_approach_normal_path}
\end{figure}

Figures \ref{fig_approach_path_failure} and \ref{fig_approach_normal_path} 
show examples of data-flow models in the form of path constraints for a failure path and a normally terminated path, respectively. 
The constraints are shown in the KQuery format used by the KLEE symbolic engine. The formula {\tt Eq 1 (ReadLSB w32 0 isEncryptedRes)} corresponds to the atomic constraint $\mathit{isEncryptedRes} = 1$.

Path constraints typically represent control-flow decisions with regard to symbolic inputs. Here, however, we collect and record certain data values in the path constraints. To do this we add instrumentation in the code that is relevant to checking feature constraints 
and introduce some symbolic metadata variables. 

\begin{figure}[th!]
\centering
\begin{footnotesize}
\begin{verbatim}
 void VerifyForward_spec__1(int client , int msg ) 
{ int pubkey ;
  int tmp, tmp___0, tmp___1 ;
  puts("before deliver\n");
  tmp___1 = isVerified(msg);
  if (tmp___1) {
    tmp = getEmailFrom(msg);
    tmp___0 = findPublicKey(client, tmp);
    pubkey = tmp___0;
    if (pubkey == 0) 
      __automaton_fail();
  }
  return;
}           
\end{verbatim}
\end{footnotesize}
\caption{An Example Code that Checks the Specification for the Unwanted 
Feature Interaction between the {\tt Verify} and {\tt Forward} features.}
\label{fig:spec}
\end{figure}   

Specifically, we identify and annotate functions that return boolean or integer values, as such functions are commonly used to check certain conditions and may be used in feature constraint checking, as shown in Figure \ref{fig:spec}. There, the function {\tt VerifyForward\_spec\_\_1} checks 
if the email has been verified and, if so, asserts that the 
public key is available for encryption before the message is forwarded.

Figure \ref{fig:isver} shows the annotated version of the {\tt isVerified} function from the Email product line. First, we introduce a symbolic metadata variable, {\tt isVerifiedRes} (line 1) and make it symbolic using the {\tt klee\_make\_symbolic} intrinsic function (line 4). Then we constrain the current symbolic execution path with a generic equality predicate 
{\tt metadata variable == returned value} using the {\tt klee\_assume} intrinsic function (lines 9, 14, and 18). This annotation does not have any side-effect on the program semantics as our annotation does not change any 
variables in the original program.  The annotation does not affect the control-flow since the added expression is consistent with the value passed to the return value. Also, each time the function is called, a new instance of the symbolic metadata variable will be created. This ensures that the values returned at different callsites do not interfere to change the control-flow semantics. Figure \ref{fig_approach_path_failure} gives an example of the multiple instances of these 
symbolic metadata variables. There  {\tt findPublicKeyRes} ends up having two instances/versions: {\tt findPublicKeyRes\_1} and {\tt findPublicKeyRes\_2}.

\begin{figure}[th!]
\centering
\begin{footnotesize}
\begin{verbatim}
int isVerifiedRes;
int isVerified(int handle ) 
{ int retValue_acc ;
  klee_make_symbolic(&isVerifiedRes,sizeof(int),
                                  "isVerifiedRes"); 
 {
  if (handle == 1) {
    retValue_acc = ste_email_isSignatureVerified0;
    klee_assume(isVerifiedRes == retValue_acc);
    return (retValue_acc);
  } else {
    if (handle == 2) {
      retValue_acc = ste_email_isSignatureVerified1;
      klee_assume(isVerifiedRes == retValue_acc);
      return (retValue_acc);
    } else {
      retValue_acc = 0;
      klee_assume(isVerifiedRes == retValue_acc);
      return (retValue_acc);
    }
  }
  klee_assume(isVerifiedRes == retValue_acc);
  return (retValue_acc);
}
}
\end{verbatim}
\end{footnotesize}
\caption{An Annotated Version of the {\tt isVerified} Function that Saves a Symbolic Constraint about the Return Value Using the Metadata Variable {\tt isVerifiedRes}.}
\label{fig:isver}
\end{figure}
Finally, two examples of data-flow dependency between feature-relevant instructions are given in Figure \ref{fig:ssexample}. They appear in lines 1173 and 1181 (line numbers are inserted as comments here) of coreutils/ls.c in BusyBox 1.32.0.
There is a feature-relevant Store-Load dependency and a Store-Store dependency between lines 1173 and 1181. This is because there are store instructions 
accessing the {\tt option\_mask32} variable at lines 1173 and 1181, and 
there is a load instruction accessing the same variable in line 1181.

\begin{figure}[th!]
\centering
\begin{footnotesize}
\begin{verbatim}
if (ENABLE_FEATURE_LS_RECURSIVE && (opt & OPT_d))
   option_mask32 &= ~OPT_R; // Line 1173    
if (!(opt & OPT_l)) { /* not -l? */
   if (ENABLE_FEATURE_LS_TIMESTAMPS && ENABLE_FEATURE_LS_SORTFILE)
      if (opt & (OPT_c|OPT_u)) {
         option_mask32 |= OPT_t; // Line 1181
/* choose a display format if one was not already specified by an option */
if (!(option_mask32 & (OPT_l|OPT_1|OPT_x|OPT_C))) // Line 1187
                option_mask32 |= (isatty(STDOUT_FILENO) ? OPT_C : OPT_1); 
\end{verbatim}
\end{footnotesize}
\caption{A Store-Store and a Store-Load Dependency between  {\tt ENABLE\_FEATURE\_LS\_RECURSIVE} and 
the  Configuration Options {\tt ENABLE\_FEATURE\_LS\_TIMESTAMPS} and 
{\tt ENABLE\_FEATURE\_LS\_SORTFILE} within coreutils/ls.c in BusyBox 1.32.0.}
\label{fig:ssexample}
\end{figure}

\section{Learning Feature Interactions}
\label{sec:learning}

\begin{figure*}[th!]
  \centering
  \scalebox{0.8}{\includegraphics[width=\linewidth]{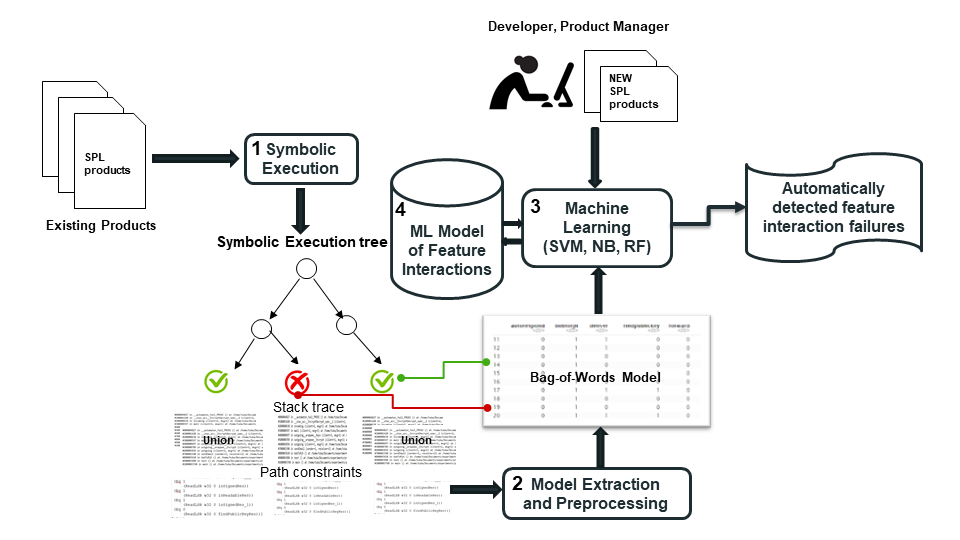}}
 
  \caption{Proposed Framework to Detect Unwanted Feature Interactions in a Software Product Line}
  \label{fig:framework}
\end{figure*}

In this section we describe the process by which we aim to learn unwanted feature interactions. In Section \ref{sec:learningspecs} we describe how our FINCH approach learns unwanted feature interactions  whenever specifications of constraints on permissible feature combinations exist for a software system or product line.  In Section \ref{sec:learningnospecs} we describe how GOLDFINCH, our generalized FINCH approach, can detect potential, unwanted feature-relevant dependencies even when  feature-constraint specifications are not available. 

\subsection{Learning unwanted feature interactions with specifications}
\label{sec:learningspecs}

Figure \ref{fig:framework} shows the steps FINCH takes to predict unwanted feature interactions when  specifications of feature-constraints are available for it to use. 

In Step 1, the implementations of the existing products, here C code, is input to the symbolic execution engine to extract the stack traces and path constraints related to each product in our repository. Stack traces contain the functions called, and path constraints contain constraints along a path. Figures \ref{fig_approach_stack_failure}, \ref{fig_approach_path_failure}, \ref{fig_approach_stack_normal} and \ref{fig_approach_normal_path} have shown the stack trace and the path constraints for both a normal and an unwanted feature interaction failure path, respectively. In Step 2, we automatically preprocess, clean and extract the functions that are called as well as the constraints from the files. In Step 3 we build a bag-of-words model
\cite{aggarwal2012survey}  used as input to the learning algorithms. Therefore, we have two different sources of data, functions in the stack trace and atomic constraints from the path condition.  We also investigate a combination of stack trace and path constraints data, which we will refer to as {\em combined data}. In Step 4 the saved learning models are used to answer developers' and product-line engineers' queries about possible unwanted feature interactions in new software product-line products.

We use supervised learning to learn from existing failure paths and success paths related to unwanted feature interactions and label a future path. When a developer combines a set of features to build a new product, the classifier helps developers know early on if this combination of features produces unwanted feature interactions in a new product. We describe in detail the use of our learning algorithms to classify the bag-of-word models in Section \ref{sec:results} below.

\subsection{Learning feature interactions without specifications}
\label{sec:learningnospecs}

\begin{figure*}[th!]
  \centering
   \scalebox{0.8}{ \includegraphics[width=\linewidth]{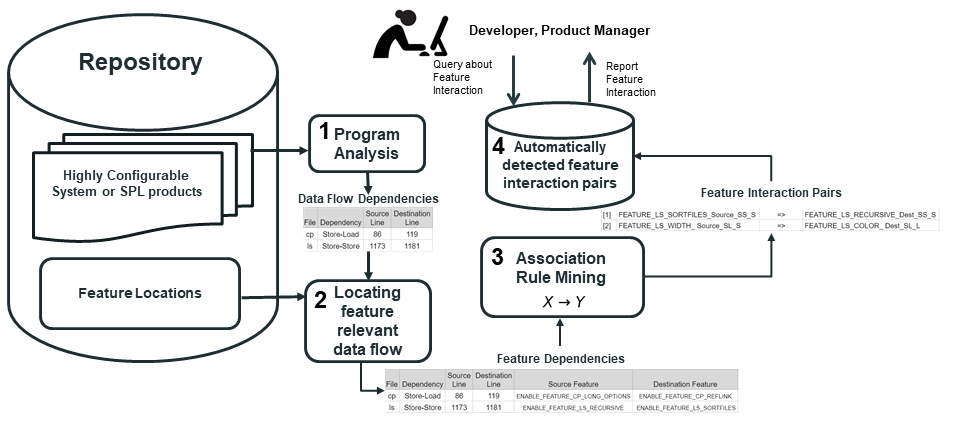}}
  \caption{Proposed Framework to Detect Unwanted Feature Interactions in a Highly Configurable System without Specifications}
  \label{fig:framework_withou_specs}
\end{figure*}

 What if there are not any specifications of  constraints on the feature combinations, or if the specifications of those constraints are partial, outdated or inaccurate?  This often happens in practice for both long-lived software product lines and configurable systems. For example, Fischer et al. reported in 2018 that it was difficult for them to find variability models that matched the implementations \cite{fischer2018predicting}. 

This lack of feature-constraint specifications seriously complicates detection of unwanted feature interactions in a new product or configuration.

GOLDFINCH offers an automated way to learn unwanted feature interactions even in the absence of feature-constraint specifications.

Figure \ref{fig:framework_withou_specs} shows how GOLDFINCH detects potential unwanted feature interactions when we do not have any feature-constraint specifications. Steps 1 and 2 were described in Section \ref{sec:modex}.  

Features in a C preprocessor-based highly configurable system are defined in the code by \#ifdef blocks \cite{hunsen2016preprocessor}.  
We use the 
cppstats tool \cite{cppstats}, which measures preprocesor-based variability, to automatically locate features and store this information in the repository, as shown in Figure \ref{fig:framework_withou_specs}. 

 In Step 1, we apply our program analysis tool PROMPT on C preprocessor-based code of a highly configurable system \cite{yavuz2020analyzing}. PROMPT performs component-level symbolic execution and supports various types of environment modeling \cite{yavuz2020tutorial}. We here extended PROMPT  to output the data-flow dependencies of the C preprocessor-based code of a highly configurable system as explained in Section \ref{sec:extraction}. We focus on two types of memory dependencies, \textbf{Store-Load} and \textbf{Store-Store}, obtaining  pairs of instructions that access the same memory location. 
 
 We use these data-flow dependencies \cite{rhein2018variability} as a useful source of data toward detecting potential unwanted feature interactions.  This is because in many cases a \textbf{Feature Dependency} occurs, i.e., one feature accesses a memory location to read and write data that another feature has already changed.

\begin{figure}[!h]
\centering
\scalebox{1}{
\includegraphics[width=3.5in]{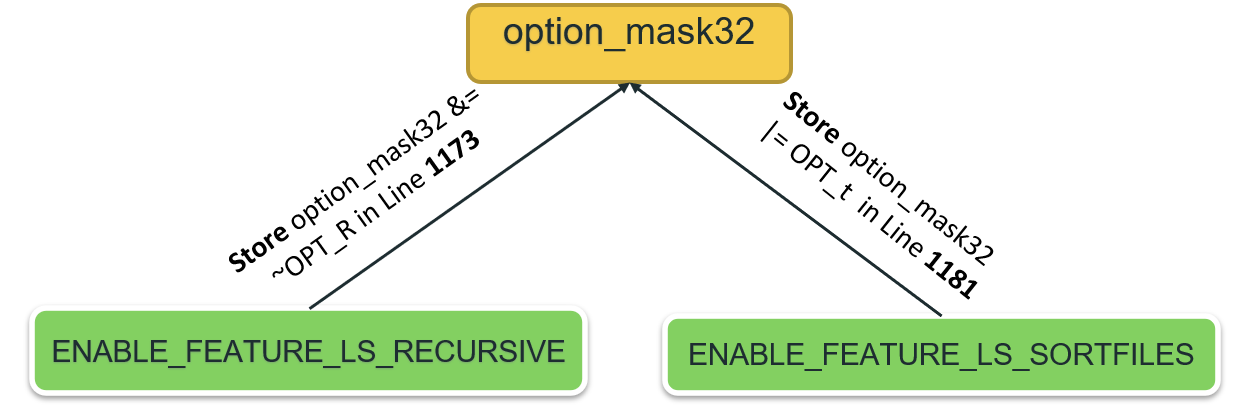}}
\caption{An Explanation of a Store-Store Data Flow Dependency that Occurs in the ls.c File of coreutils in BusyBox 1.32.0.}
\label{fig_source-destination_feature}
\end{figure}

  \begin{figure}[th!]
  \centering
  \includegraphics[width=\linewidth]{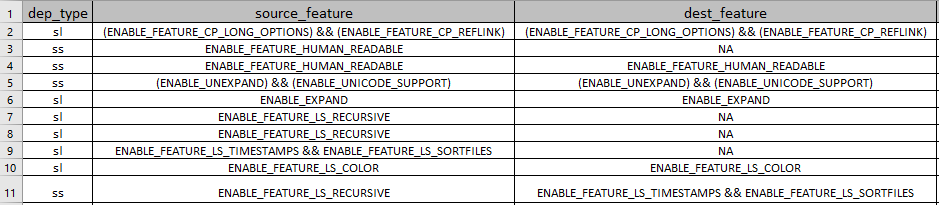}
  \caption{A Snapshot of Feature Dependencies in coreutils of BusyBox 1.32.0}
  \label{fig:feature_dependencies}
\end{figure}
In Step 2, GOLDFINCH uses this feature location data to automatically  relate each line of a data-flow dependency with its corresponding features. For example, Figure \ref{fig_source-destination_feature} displays a Store-Store data-flow dependency that occurs in line 1173 and 1181 of the ls.c file of coreutils of BusyBox 1.32.0.  In Step 1, PROMPT outputs this pair of lines. In Step 2, we automatically find the corresponding features related to these lines, labeling them as \textbf{Source-Features} and \textbf{Destination-Features}. In this same way, we obtain all feature-dependency pairs of our  configurable system. 
 
  In a highly configurable system, there may be many feature dependency records. 
 In Step 3, we learn from the large set of feature dependency pairs extracted in Step 2. 
 To automate learning from the existing feature dependency records, we use frequent item set mining, i.e., association rule mining, to learn the most frequent dependent features. We apply the Apriori algorithm \cite{agrawal1994fast}, an unsupervised learning method for frequent item set mining and association rule learning over relational databases. Association rule mining enables us to learn which pairs of features occur together based on the large amount of data we have \cite{borgelt2012frequent,rhein2018variability}.

 The Apriori algorithm uses parameters “support” and “confidence”. Support is the frequency of an item in a dataset. Confidence is how likely two items are to occur together in the dataset. Each record in our use of the Apriori algorithm is a set of two items. An item is a feature, and an item set refers to features that are dependent and control the same memory region. For example, one input to the Apriori algorithm is an item set of two features {\tt ENABLE\_FEATURE\_LS\_RECURSIVE} and {\tt ENABLE\_FEATURE\_LS\_SORTFILE} that control the variable {\tt option\_mask32} and are dependent. In some cases, ``OR", ``AND" or ``NOT" combination of features control a memory region. We consider the whole logical expression as an item. For example, the logical expression of  {\tt ENABLE\_UNEXPAND} \&\& {\tt ENABLE\_UNICODE\_SUPPORT} controls a memory region and thus is considered an item. 
 
\begin{figure}[th!]
  \centering
  \scalebox{0.6}{
  \includegraphics[width=\linewidth]{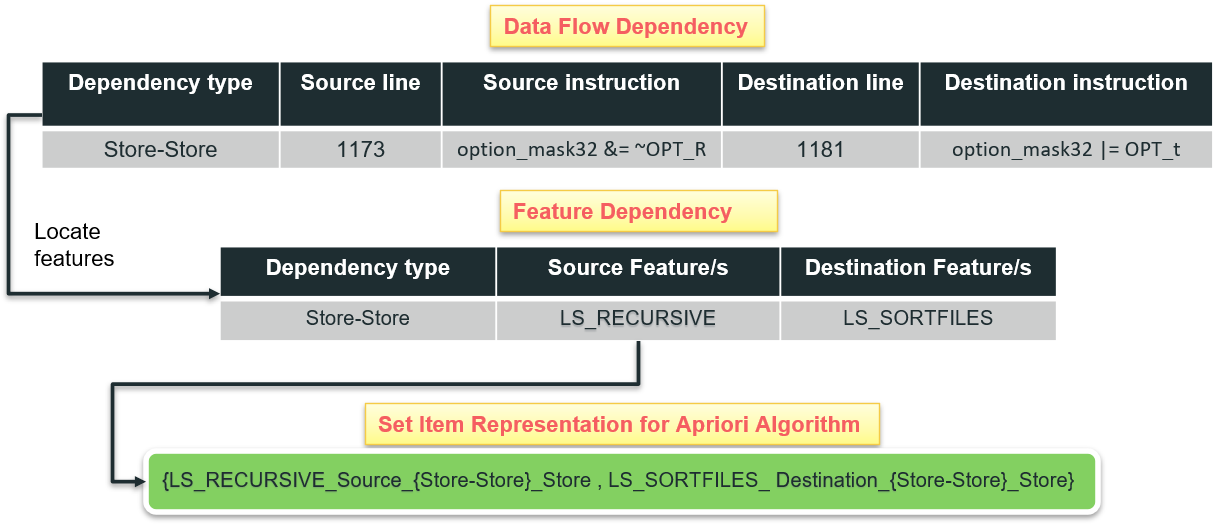}}
  \caption{Set Item Representation of Feature Dependency for Unsupervised Learning}
  \label{fig:set_rep}
\end{figure}

 Figure \ref{fig:feature_dependencies} shows a snapshot of feature dependencies in coreutils of BusyBox 1.32.0. Figure \ref{fig:set_rep} shows how we convert the feature dependencies shown in Figure \ref{fig:feature_dependencies} to an item set in order to feed it into the Apriori algorithm.

{\it Encoding.} The input to the Apriori algorithm is a set representation of items, here the features, used to find the relevant items. The feature-dependency pairs thus must be encoded with the information as to whether the feature is a  source feature or a destination feature, as well as the data-flow dependency type.  We encode each feature dependency in the frequent item set as follows: (1) each feature can be related to the source or the destination in each data flow dependency; (2) each item is labeled with data-flow dependency type; and (3) each feature controls the source or destination part of the data-flow dependency.

Figure \ref{fig:set_rep} gives an explanatory example of the process of encoding a data-flow dependency into a set representation of feature dependency appropriate for input to the Apriori algorithm. It shows the existence of  a 
data-flow dependency in lines 1173 and 1181. The data-flow dependency type is Store-Store, meaning that both lines write to the same variable, namely option\_mask32. Line 1173 is in the scope of  
feature LS\_RECURSIVE. The memory access type is a Store (S).  
We thus encode this feature as $LS\_RECURSIVE\_Source\_\{Store-Store\}\_Store$ for input to the Apriori algorithm.

In this way, we can encode the dependency type, source/destination, and the specific feature combination expression into the set representation used by the Apriori algorithm to learn the features most dependent on each other. 
 
 The output of Step 3 is a set of association rules showing features that are dependent on each other. For example, the following association rule shows that the two features $LS\_RECURSIVE$ and $LS\_SORTFILE$ are dependent on each other: \newline

$LS\_RECURSIVE\_Source\_\{Store\_Store\}\_Store \iff$ \\
$LS\_SORTFILE\_Destination\_\{Store\_Store\}\_Store$. \\

Not all frequent item sets detected will be of interest. We thus filter out for developers' consideration only the pairwise feature set where the source and destination feature differ. That is, if the Source-Feature and Destination-Feature in a pair are identical, we exclude it, as shown in Figure \ref{fig:feature_dependencies} record number 10.  

In Step 4, the final filtered pairs are stored, and GOLDFINCH reports the association rules to the developers, since these identify feature interactions 
to be checked for whether they are unwanted.
The intended usage scenario is for results to be made available to developers, either through system-wide reports or through queries regarding  specific feature combinations of concern. 

  \begin{table*}[th!]
   \caption{Description of the three Software Product Lines}
 \label{tab:case_studies}
  \centering
  \begin{footnotesize}
  \begin{tabular}{|l|c|c|c|c|c|c|}
  \hline
  SPL name & \# Features & \# Products & \# Feature  & \# Normal  & \# Unwanted Feature & \% Failure \\
  
   &  &  & Interaction &  path &  Interaction path &  Ratio\\
  
  \hline
  \hline
  Email \cite{apel2013feature} & 9 &36 &10  &13595 & 101 & 0.7\\
  Elevator \cite{apel2013feature} & 6 & 20  & 5 & 84 & 49 & 36.9 \\
  Mine Pump \cite{apel2013feature} & 7 &64 & 5 &27775 & 96 & 0.34\\
 \hline
  \end{tabular}
  \end{footnotesize}
 \end{table*}

\begin{table}[th!]
\caption{Features in the Email Software Product Line, extracted from \cite{apel2011detection}}
\begin{center}
\begin{footnotesize}
\begin{tabular}{|c|c|}
\rowcolor{lightgray}
\hline
Feature name & Short description\\
\hline
Addressbook & manage e-mail contacts
\\
\hline
Autoresponder & respond to e-mails\\
\hline
EmailClient or Base & basic e-mail client\\
\hline
Decrypt & decrypt incoming e-mails\\
\hline
Encrypt & encrypt outgoing e-mails\\
\hline
Forward & forward incoming e-mails\\
\hline
Sign & sign outgoing e-mails\\
\hline
Verify & verify e-mail signatures\\
\hline
\end{tabular}
\end{footnotesize}
\label{tab1}
\end{center}
\end{table}

\begin{table}[th!]
\caption{Known Feature Interactions, with their Identifiers from the Original Sources \cite{hall2005fundamental,apel2011detection}}
\begin{center}
\begin{footnotesize}
\begin{tabular}{|c|c|}
\hline
\rowcolor{lightgray}
Feature Interaction Id & Features Involved\\
\hline
0 & Decrypt, Forward\\
\hline
1 & Addressbook, Encrypt\\
\hline
3 & Sign, Verify\\
\hline
4 &Sign, Forward\\
\hline
6 & Encrypt, Decrypt\\
\hline
7 & Encrypt, Verify\\
\hline
8 & Encrypt, Autoresponder\\
\hline
9 & Encrypt, Forward\\
\hline
11 & Decrypt, Autoresponder\\
\hline
27 & Verify, Forward\\
\hline
\end{tabular}
\end{footnotesize}
\label{tab:known_fi_email}
\end{center}
\end{table}

In summary, we use the Apriori algorithm to learn which features depend upon each other, either in a highly configurable system or in a software product line's systems. This enables automated detection of feature dependencies indicative of possible unwanted feature interactions even when no feature-constraint specifications exist.

\subsection{Case Studies}

Table \ref{tab:case_studies} shows the three software product lines used in our evaluation: the Email product line, 

\cite{hall2005fundamental}, the Elevator product line \cite{plath2001feature}, and the Mine Pump product line \cite{kramer1983conic}.  All three have been used as benchmarks in the software product line literature \cite{apel2011detection, apel2013strategies} and have a variety of features with potential, unwanted interactions causing failed executions if they occur. We use the software product line versions written in C and provided in \cite{apel2013strategies}.

As an example, we look more closely at the Email software product line. Table \ref{tab1}   shows the eight available features (units of functionality) in it. The base feature is Email Client, which is shared by all products.  There are ten pairwise known unwanted feature interactions as shown in Table  \ref{tab:known_fi_email}. 
By ``known unwanted feature interactions'', we mean that they are documented in the product line's feature specifications.  We refer the reader to \cite{colder2000feature} and \cite{hall2005fundamental} for details of the other interactions shown in the table.  The Elevator product line similarly has six features and five unwanted feature interactions specified, and the Mine Pump product line has seven features and five unwanted feature interactions specified.  

We also investigated and refined our approach through application on a large, highly configurable system. BusyBox is a  collection of many Unix utilities implemented in a single executable file. We used the latest version of BusyBox 1\_32\_0, and

focused on the \textbf{coreutils} section of BusyBox, as it had the most problematic interactions based on this study \cite{abal2018variability}. 
Coreutils of BusyBox 1\_32\_0 has 139 features and about 19k lines of code. With 139 features, we can have 9591 pairwise options, making it  

a very large configurable system \cite{thianniwet2016scaling}.

\section{Evaluation and Results}
\label{sec:results}

In this section we describe the results from our investigation of the following research questions in four case studies. We first evaluate FINCH's discovery of unwanted feature interactions when {\it feature-constraint specifications are  available}, as in many product lines. We then evaluate GOLDFINCH's discovery of feature interactions when such {\it specifications are not available}, as in most highly configurable systems.  The key research questions are the following.

\begin{itemize}
\item \emph {RQ1:  How effective is the proposed approach in producing accurate results for known feature interactions?}

\item \emph{RQ2:  How accurately can the proposed approach predict new unwanted feature interactions based on existing feature interactions?}  

\item \emph{RQ3:  How scalable is the performance of our approach when specifications are unavailable?}

\end{itemize}

\subsection{Detecting unwanted feature interactions with specifications }
\label{sec:eval_with}

In this subsection we  describe evaluation results for systems where feature-constraint specifications exist, that is, when we can use FINCH. 
To answer the research questions for FINCH, we first built a bag-of-words model \cite{zhang2010understanding,aggarwal2012survey} of the execution traces and path constraints extracted, using our extension to the KLEE 
symbolic execution engine \cite{cadar2008klee}, for the Email, Elevator, and Mine Pump software product lines \cite{apel2013feature} described in Table \ref{tab:case_studies}. We used the following setup for our experiments: 

\begin{itemize}
    \item Balanced Accuracy, defined as
    (BAC = 0.5*(True Positive/(True Positive + False Negative) + True Negative/(True Negative + False Positive)), the average accuracy obtained on two classes, is used for reporting the accuracy of the machine-learning models. \cite{brodersen2010balanced}.
   \item 80\% of the data is used for the training of the models and 20\% for the testing.
   \item 5 repeats of 10-Fold Cross Validation are used inside the training.
    \item SMOTE sampling is used inside the training since the data is imbalanced, with the number of records with a failure label being very small compared to the number of records with a normal label, as shown in Table \ref{tab:case_studies} \cite{chawla2002smote}.
\end{itemize}

\textit {\textbf{RQ1.F:} How effective is FINCH in classifying paths in new products for known feature interactions?}

This research question investigates whether we can 
classify termination paths into two classes, normal termination (success) and interaction-associated failed termination (failure). 
The ``F'' suffix in the RQ indicates that its investigation uses FINCH.  We selected three well-regarded machine learning classifiers to categorize the text  \cite{joachims1998text,aggarwal2012survey}: Support Vector Machine (SVM), Naive Bayes, and  Random Forest.  We learned from the existing labeled data, with the labels being failure path and success path, built the classifier, and classified previously unseen traces of functions and atomic constraints from the path conditions.

\begin{figure*}[!h]
\centering
\begin{subfigure}[b]{0.3\textwidth}
\includegraphics[width=\textwidth]{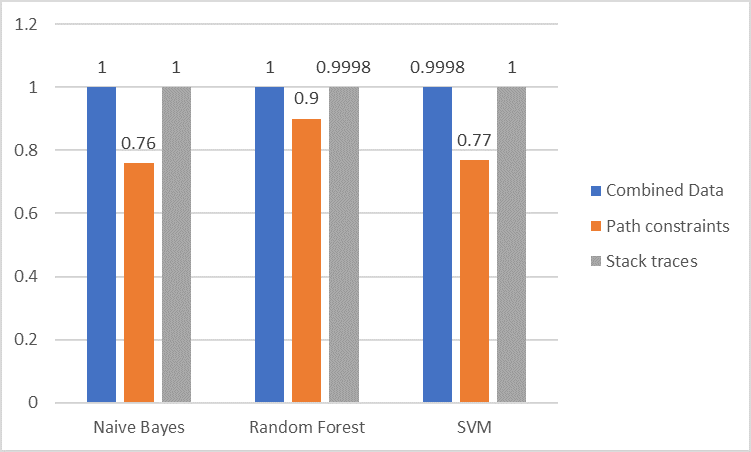}
\caption{Balanced Accuracy}
\end{subfigure}
\begin{subfigure}[b]{0.33\textwidth}
\includegraphics[width=\textwidth]{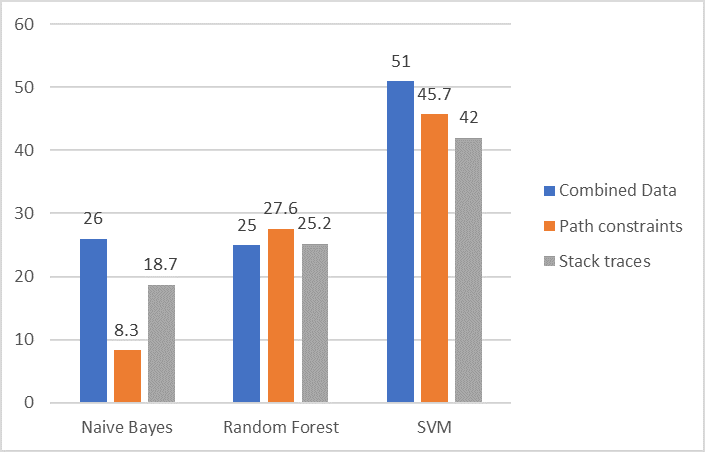}
\caption{Training Time}
\end{subfigure}
\begin{subfigure}[b]{0.33\textwidth}
\includegraphics[width=\textwidth]{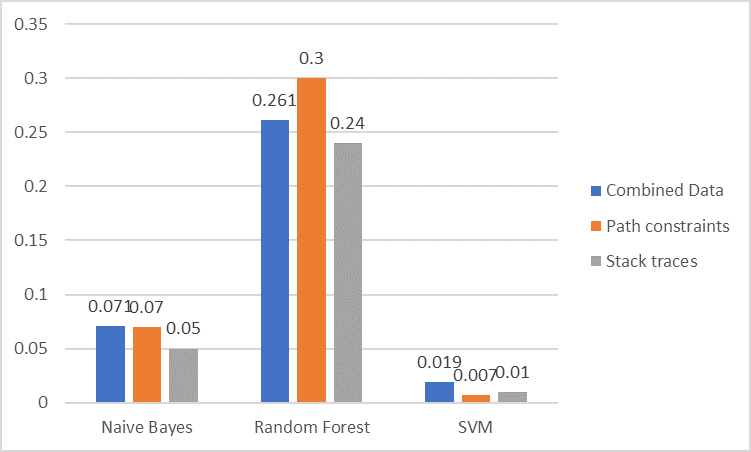}
\caption{Prediction Time}
\end{subfigure}
\caption{Email: Performance and Time Evaluations}
\label{fig:rq1_email}
\end{figure*}

\begin{figure*}[!h]
\centering
\begin{subfigure}[b]{0.33\textwidth}
\includegraphics[width=\textwidth]{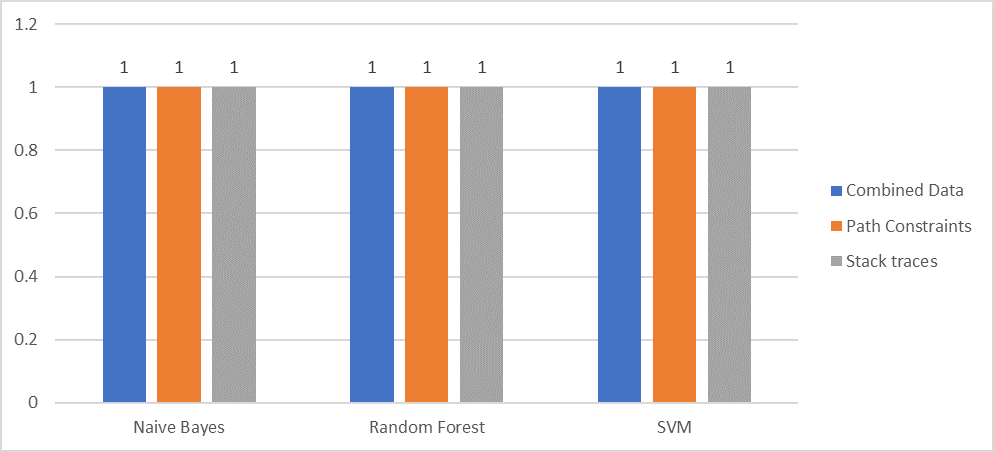}
\caption{Balanced Accuracy}
\end{subfigure}
\begin{subfigure}[b]{0.33\textwidth}
\includegraphics[width=\textwidth]{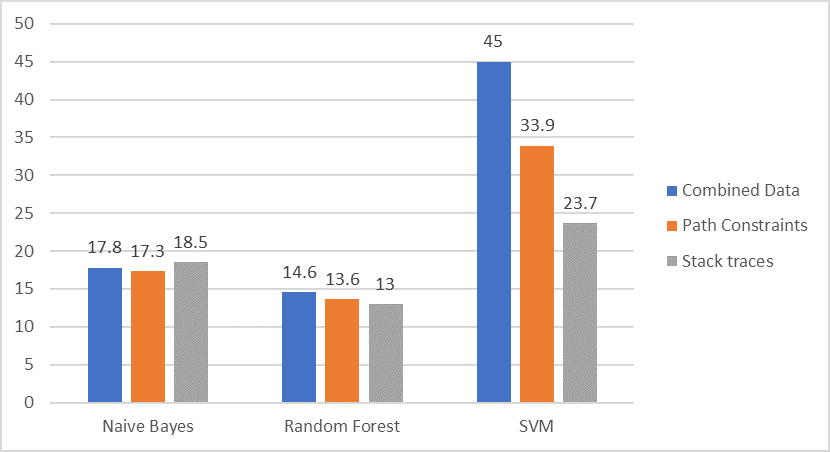}
\caption{Training Time}
\end{subfigure}
\begin{subfigure}[b]{0.33\textwidth}
\includegraphics[width=\textwidth]{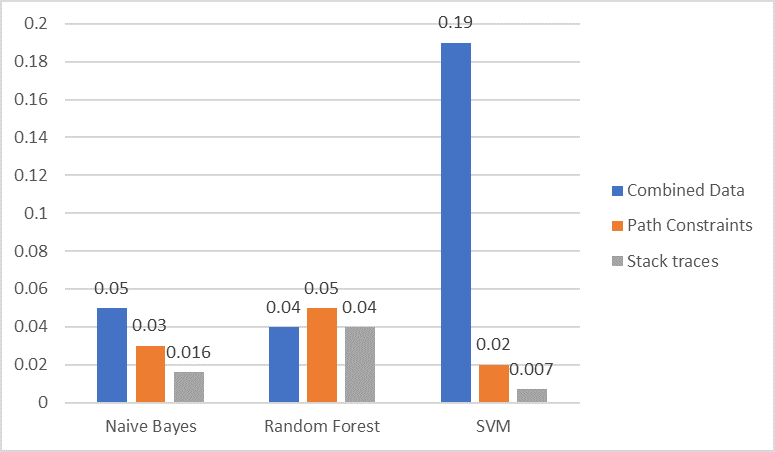}
\caption{Prediction Time}
\end{subfigure}
\caption{Elevator: Performance and Time Evaluations}
\label{fig:rq1_elevator}
\end{figure*}

\begin{figure*}[!h]
\centering
\begin{subfigure}[b]{0.33\textwidth}
\includegraphics[width=\textwidth]{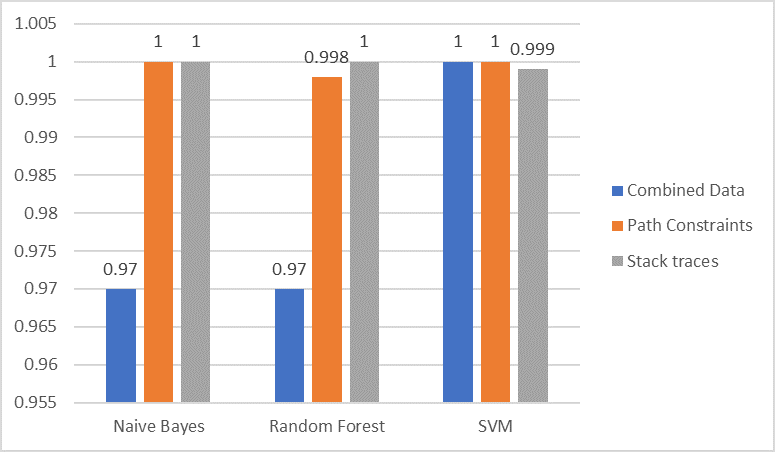}
\caption{Balanced Accuracy}
\end{subfigure}
\begin{subfigure}[b]{0.33\textwidth}
\includegraphics[width=\textwidth]{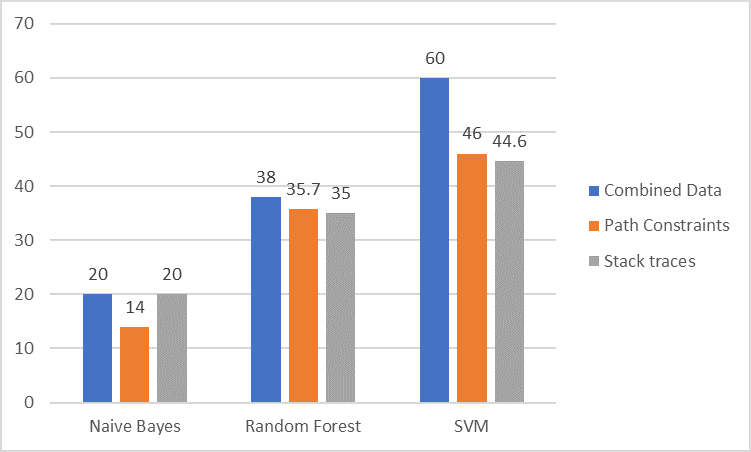}
\caption{Training Time}
\end{subfigure}
\begin{subfigure}[b]{0.33\textwidth}
\includegraphics[width=\textwidth]{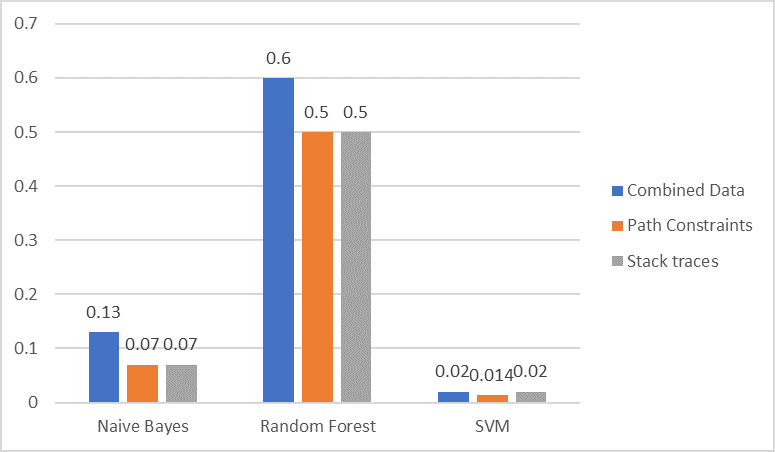}
\caption{Prediction Time}
\end{subfigure}
\caption{Mine Pump: Performance and Time Evaluations}
\label{fig:rq1_minepump}
\end{figure*}

The feature vector produced by our bag-of-words model, as shown in Figure \ref{fig:framework}, fed into these classifiers. We divided the data to perform 10-fold cross-validation and reported the corresponding Balanced Accuracy, Training Time, and Prediction Time values. Figures \ref{fig:rq1_email}, \ref{fig:rq1_elevator} and \ref{fig:rq1_minepump} show these values for the three software product lines. 

We experimentally tested the three selected ML algorithms on three different sources of data: (1) functions that appear on the stack traces, (2) path constraints, and (3) combinations of both stack traces and path constraints. For each case study, we investigated how accurately we were able to classify failure and normal cases. We also investigated which of these sources of data aided the classifier in distinguishing between the normal and failure classes for a path.

Results shown in Figure \ref{fig:rq1_email} for the Email product line indicate that our approach was able to correctly detect failure cases and could accurately classify the data using just the stack trace data. For the Elevator product line, as shown in Figure \ref{fig:rq1_elevator}, we could perfectly classify the failure and normal termination paths using either path constraints or stack traces.  In Figure \ref{fig:rq1_minepump}, the results on the Mine Pump data show that the path constraints data led to a slightly more accurate model for SVM. 

In summary, our experiments showed that all three learned models had high Balanced Accuracy. In Mine Pump, SVM surpassed Naive Bayes and Random Forest.  Using both path constraints and stack traces yielded an accurate model for all three software product-line case studies.  

To investigate why there was a more accurate model if we used stack traces for the Email product line, and a more accurate model if we used path constraints for the Mine Pump product line, we measured how much of the data for each was provided from stack traces and how much from path constraints.  Table \ref{tab:rq1_length} describes both sources of data in terms of the length of the minimum path:maximum path for each product line's unique functions, and in terms of the min:max size of its path constraints. 

As shown in Table \ref{tab:rq1_length}, there are 33 unique functions in the stack traces of the Email product line and 11 unique constraints in its path constraints. This indicates that we had more data in the stack traces compared to the path constraints. This appears to lead to a more accurate model using stack traces for the Email product line.  For the Mine Pump product line, we speculate that its larger min-to-max range of path lengths for the Path Constraints Data assists in learning a more accurate model using path constraints.  However, for the Elevator product line, which also has a large min-to-max path range,  learning the model with both sources of data yielded a perfect classifier.  This may be due to its having a sufficient number of both unique functions and unique path constraints.  This leads us to {\bf recommend learning a model on both stack-trace data and path-constraint data}. It further suggests that study of what constitutes sufficient stack-trace and path-constraint data for learning a product line model would be useful.  

Regarding the computation time, SVM took longer to build a model in all three case studies; however, SVM was also the fastest algorithm in predicting an unwanted feature interaction.  Overall, SVM displayed better performance compared to Naive Bayes and Random Forest in our experiments.

{\bf Finding: Our approach accurately classified failure paths related to unwanted feature interactions and identified all known unwanted feature interactions for the three product lines.}

To try to answer RQ2, ``How accurately can the proposed approach predict new unwanted feature interactions based on existing feature interactions?,"  we investigated three sub-questions, RQ2.1F, RQ2.2F, and RQ2.3F.  We discuss each of these below.

\textit {\textbf{RQ2.1F:}  Can FINCH identify a new unwanted feature interaction?}

 Recent studies show that a new feature similar to an existing feature involved in an existing feature interaction tends to behave similarly \cite{khoshmanesh2018role,khoshmanesh2019leveraging,9233030}. Therefore, we want to know if having information about existing unwanted feature-interaction paths helps predict new unwanted feature interactions.

To evaluate our classifiers on new unwanted feature interactions, we performed the following steps: 
1) select a feature interaction pair; 2) exclude the data related to this feature interaction; 3) build the classifiers on the remaining data with 10-fold cross-validation; 4) test the classifiers on the new data excluded in step 2; and 5) repeat for all 10 feature interaction pairs in the Email benchmark.

\begin{table}[th!]
  \centering
  \caption{Comparing Path Lengths and Unique Functions and Constraints for Each Software Product Line}
    \begin{footnotesize}
    \begin{tabular}{|l|r|r|r|r|}
    \toprule
          & \multicolumn{2}{c|}{\textbf{Stack Trace Data}} & \multicolumn{2}{c|}{\textbf{Path Constraints Data}} \\
    \midrule
    \textbf{SPL name} & \multicolumn{1}{l|}{\textbf{Path length(min:max)}} & \multicolumn{1}{l|}{\textbf{\#Funcs.}} & \multicolumn{1}{l|}{\textbf{Const. size(min:max)}} & \multicolumn{1}{l|}{\textbf{\# Const.}} \\
    \midrule
    \textbf{Email} & 5:18  & 33    & 1:25  & 11 \\
    \midrule
    \textbf{Elevator} & 2:16  & 24    &28:223 & 37 \\
    \midrule
    \textbf{Mine Pump} & 4:15  & 21    & 3:49  & 15 \\
    \bottomrule
    \end{tabular}
    \end{footnotesize}
  \label{tab:rq1_length}%
\end{table}%

\begin{table}[th!]
  \centering
  \caption{Percentage of Detected Unwanted Feature Interactions}
  \begin{footnotesize}
  \begin{tabular}{|l|r|r|r|}
    \toprule
          \textbf{Across all ML Algorithms} & \multicolumn{1}{l|}{\textbf{Email}} & \multicolumn{1}{l|}{\textbf{Elevator}} & \multicolumn{1}{l|}{\textbf{Mine Pump}} \\
    \midrule
    Stack traces data & 100    & 100   & 100 \\
    \midrule
    Path constraints data & 43    & 100   & 100 \\
    \midrule
    Combined data & 100   & 100   & 100 \\
    \midrule
    \midrule
     \textbf{Machine Learning Algorithms}.& \multicolumn{3}{c|}{  }\\
     \midrule
    SVM & 100   & 100   & 100 \\
    \midrule
    Naïve Bayes & 86    & 100   & 100 \\
    \midrule
    Random Forest & 86    & 100   & 100 \\
    \bottomrule
    \end{tabular}
    \end{footnotesize}
  \label{tab:rq2}%
\end{table}%

Table \ref{tab:rq2} shows the percentage of unwanted feature interactions
that we could detect for each product line using different sources of data and different machine learning algorithms.  We ran experiments using each of the two sources of data--stack traces and path constraints--as well as with both of them combined. Furthermore, we evaluated three machine learning algorithms as to their accuracy in detection of feature interaction failures. 

Table \ref{tab:rq2} shows that SVM had the best performance among the learning algorithms in our study.  The table also shows that using the combined data yielded the highest percentage of unwanted feature-interaction failures that were detected.  These data are consistent with our findings from RQ1 above.For the Email product line, using only the path constraints data was less effective than using the stack traces for detecting new unwanted feature interactions. Stack trace data or combined data resulted in high accuracy in predicting even new feature interactions. 

\vspace{1em}

 \begin{figure*}[th!]
\centering
\begin{subfigure}[b]{0.47\textwidth}
\includegraphics[width=\textwidth]{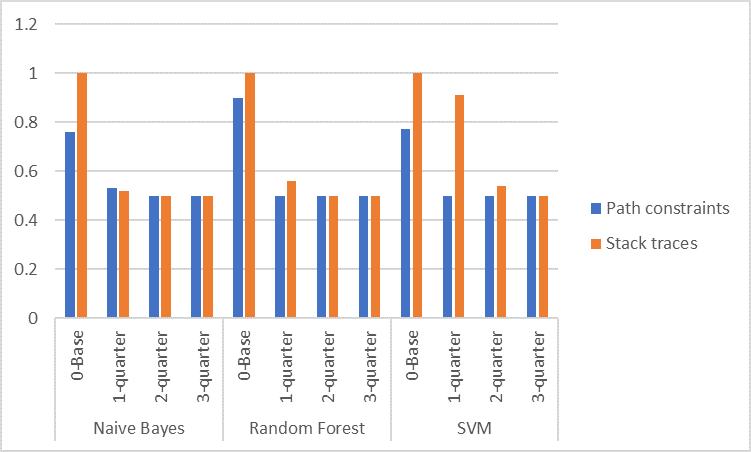}
\caption{Email}
\end{subfigure}
\begin{subfigure}[b]{0.47\textwidth}
\includegraphics[width=\textwidth]{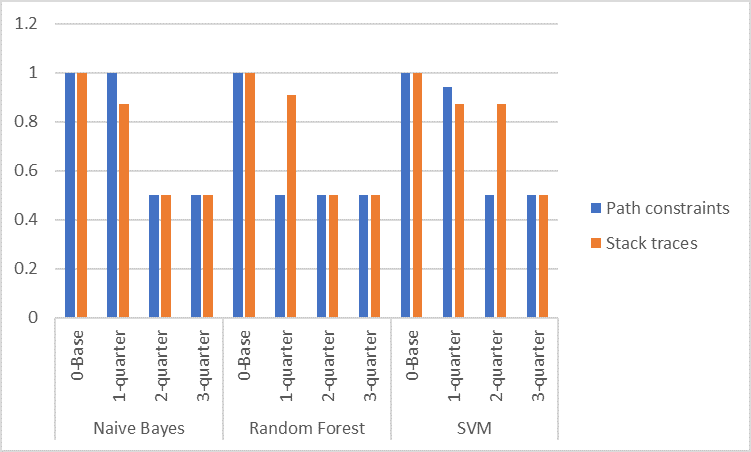}
\caption{Elevator}
\end{subfigure}
\begin{subfigure}[b]{0.5\textwidth}
\includegraphics[width=\textwidth]{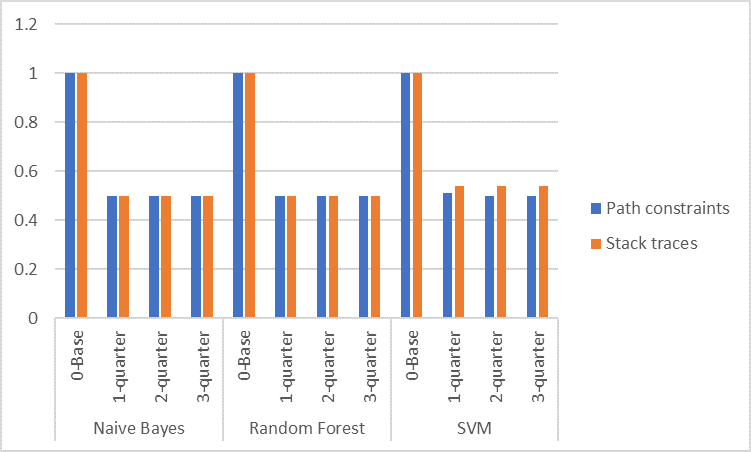}
\caption{Mine Pump}
\end{subfigure}
\caption{Balanced Accuracy vs. Partial Feature Elimination in Testing}
\label{pic:rq3}
\end{figure*}

\textit {\textbf{RQ2.2F:} How effective is FINCH in dealing with partial data in testing?}\\

 This research question asks whether we can detect new feature interaction failures with partial and unseen data. This is of interest primarily because using only partial data could speed up the computation time needed for the  testing (learning-based prediction of feature interactions) for each new product in a product line. It also is of interest because it could enable FINCH to be used earlier for a new product while only partial data is available. 
 
 To explore the accuracy of our model in dealing with unseen data, we manipulated the testing data to exclude varying portions of the features.   We used the models learned in RQ1 and tested the models as follows: 1) excluded one quarter, two quarters, or three quarters of functions and path constraints from the top of the stack traces and from the path constraints; 2) tested the models with the reduced data; and 3) checked whether and, if so, how much the accuracy of detecting failure cases deteriorates. 
 
 Figure \ref{pic:rq3} shows the balanced accuracy for each model in testing on partial data. For the Email software product line, even when one quarter of the features were excluded from the testing data, we still obtained a high balanced accuracy, 0.91, by using SVM on stack traces data in detecting failure cases.  For the  Elevator product line, the balance accuracy also was high, 0.94, when we excluded a quarter of the features on path constraints. Building the model on the stack trace data when a quarter of the data was excluded from the top of the path constraints also yielded a highly accurate model, using Naive Bayes or SVM. Even excluding two quarters of data still achieved a highly accurate model using stack trace data and SVM.

 However, in Mine Pump, we were not able to classify a path with only partial data. 
 These mixed results suggest that, for at least some product lines, our method can accurately predict many but not all unwanted feature interactions with access to only 75\% of the data related to the new product. This finding may assist with reducing the computation time needed to find unwanted feature interactions.

\textit {\textbf{RQ2.3F:} Which features are more important for FINCH?}\\

To address this research question, we worked to identify which features are more and less important for the Random Forest model described for RQ1. We used the Gini importance index to measure feature importance. Gini importance is defined as the total decrease in node impurity averaged over all trees of the ensemble model \cite{wright2015ranger}. 

 \begin{figure*}[!h]
\centering
\begin{subfigure}[b]{0.47\textwidth}
\includegraphics[width=\textwidth]{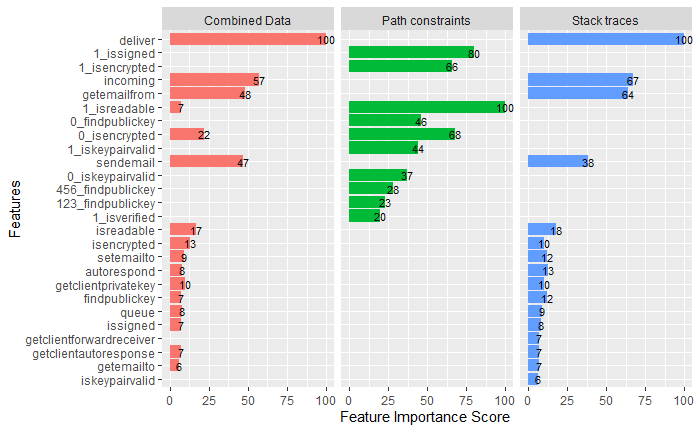}
\caption{Email}
\end{subfigure}
\begin{subfigure}[b]{0.47\textwidth}
\includegraphics[width=\textwidth]{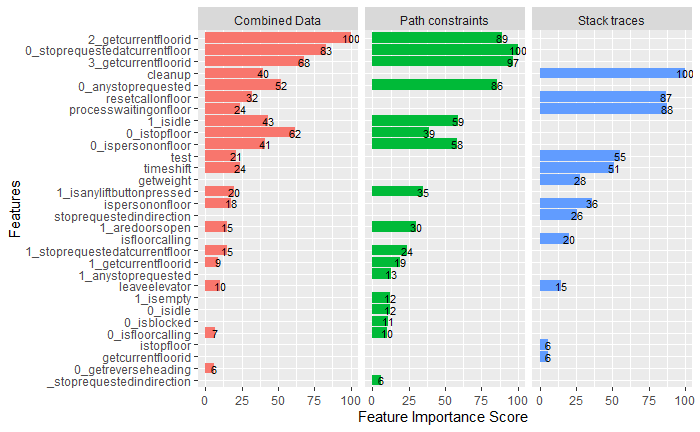}
\caption{Elevator}
\label{fig:rq4_elevator}
\end{subfigure}
\begin{subfigure}[b]{0.47\textwidth}
\includegraphics[width=\textwidth]{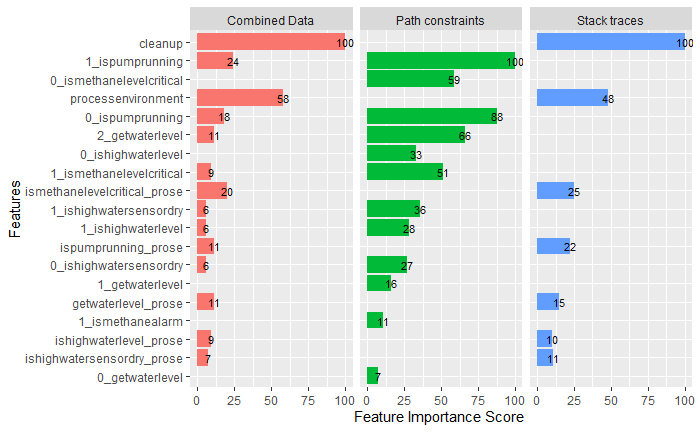}
\caption{Mine Pump}
\end{subfigure}
\begin{subfigure}[b]{0.5\textwidth}
\includegraphics[width=\textwidth]{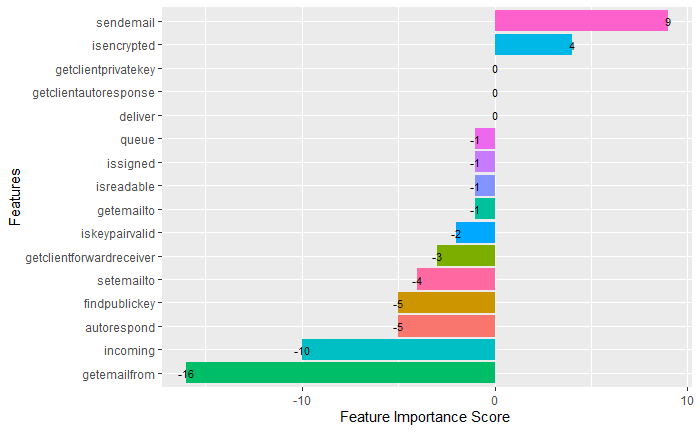}
\caption{Diff Feature Importance Scores in Email }
\label{fig:rq4_diff_email}
\end{subfigure}
\caption{Feature Importance Comparison on Stack Traces, Path Constraints, and Combined Data}
\label{fig:rq4_importance}
\end{figure*}

Figure \ref{fig:rq4_importance} shows the feature importance values for the Email, Elevator and Mine Pump product lines. The table shows how in Elevator, Figure \ref{fig:rq4_elevator}, the Random Forest model is built on the combined data, using features from both stack traces and path constraints to classify the data. This contrasts with both Email and Mine Pump where the combined model mostly uses stack trace features.  These findings indicate that the relative importance of the data captured in stack traces and path constraints may vary case by case in learning the model. This again suggests that using both sources of data--stack traces and path constraints--can result in a more accurate model for predicting unwanted feature interactions versus normal terminations.

We also can use the result of this research question to prioritize the paths and guide the symbolic execution engine to explore more problematic paths. We give an example from the Email system to show how we can take advantages of the feature importance results. Figure \ref{fig:rq4_diff_email} shows the differences in scores of feature importance for stack traces data when path constraints data are added to the model. Three functions  "deliver", "getclientautoresponse", and "getclientprivatekey" retain the same scores  when path constraints data is added while the scores of the "isencrypted" and "sendmail" functions are increasing.

\begin{table}[htbp]
  \centering
  \caption{Comparison of Learning Models' Performance on 5 Selected Features}
    \begin{footnotesize}
    \begin{tabular}{|l|c|c|c|c|c|}
    \toprule
    \multicolumn{1}{|c|}{\textbf{ML model}} & \textbf{Balanced Accuracy} & \textbf{Recall} & \textbf{Precision} & \textbf{Training Time} & \textbf{Prediction Time} \\
    \midrule
    SVM, 5 Features & 0.9974255 & 1     & 0.58  & 21    & 0.007 \\
    \midrule
    Naïve Bayes, 5 Features & 0.9974255 & 1     & 0.58  & 10    & 0.02 \\
    \midrule
    Random Forest, 5 Features & 0.9974255 & 1     & 0.58  & 22    & 0.24 \\
    \midrule
    SVM, All Features & 1     & 1     & 1     & 51    & 0.02 \\
    \midrule
    Naïve Bayes, All Features & 1     & 1     & 1     & 26    & 0.07 \\
    \midrule
    Random Forest, All Features & 1     & 1     & 1     & 25    & 0.261 \\
    \bottomrule
    \end{tabular}
    \end{footnotesize}
  \label{tab:rq4_feature_importance}%
\end{table}%

\begin{table*}[th!]
    \caption{The Data-flow Type Dependencies That Were Detected by GOLDFINCH Within the Sample Code, Explaining the Configuration Bugs in BusyBox as Reported in \cite{abal2018variability,VBDB}.}
    \label{tab:knownbugs}
    \centering
    \begin{footnotesize}
    \begin{tabular}{cccc}
    {\bf Buggy Configuration} & {\bf Bug Type} & {\bf Dependency Type} & {\bf Dependent Features} \\ \toprule
       BB\_MMU \&\& FEATURE\_HTTPD\_GZIP    & Behavior & Store-Load & ENABLE\_FEATURE\_HTTPD\_GZIP, \\
       \&\& FEATURE\_HTTPD\_BASIC\_AUTH & Violation & & FEATURE\_HTTPD\_BASIC\_AUTH \\ \hline
       BB\_FEATURE\_LS\_FILETYPES \&\&   &  Uninitialized & Store-Load & BB\_FEATURE\_LS\_FILETYPES,\\
       !BB\_FEATURE\_LS\_USERNAME & Variable & & BB\_FEATURE\_LS\_USERNAME\\ \hline
       FEATURE\_LS\_SORTFILES \&\&  & Memory & Store-Load & FEATURE\_LS\_SORTFILES, \\
       FEATURE\_LS\_RECURSIVE  & Leak & & FEATURE\_LS\_RECURSIVE \\ \hline
       FEATURE\_MDEV\_CONF \&\&  & Unused & - & - \\ 
       FEATURE\_MDEV\_RENAME  \&\& & variable & & \\    
       !FEATURE\_MDEV\_RENAME\_REGEXP  & &  &   \\
       \bottomrule
    \end{tabular}
    \end{footnotesize}
\end{table*}

Moreover, adding path constraints data helps some features from stack traces manifest themselves better in the decision-making process. We built machine-learning models using data containing the five features mentioned above to check if a model having only these features  could correctly detect all the failure cases. Table \ref{tab:rq4_feature_importance} shows the result of the machine-learning models built on these five features. SVM had the best performance and correctly detected all the failure cases (recall=1) for the 5 feature model very quickly (0.007 secs.).
This result may be important for the use of the symbolic execution engine, assisting to prioritize paths and reduce the path explosion problem.

Such reduced sets of the machine learning features can potentially be leveraged to speed up the search for detecting 
unwanted feature interactions when using exhaustive techniques such as symbolic execution. 

{\bf Finding: In experiments that excluded data related to each feature interaction, our approach predicted unseen feature interactions with high balanced accuracy.  With partial data (75\%) it predicted all but one of the unwanted feature interactions.}

\subsection{Detecting unwanted feature interaction without specifications }
\label{sec:eval_without}

We next describe our results in finding potentially unwanted feature interactions when we do not have access to specifications. 
GOLDFINCH uses the data flow dependencies described in Section \ref{sec:dfmodels} to detect unwanted feature interactions in highly configurable systems without access to specifications of  constraints on feature interactions. 

We first report results from applying GOLDFINCH on a highly configurable software system.  We then report results from applying GOLDFINCH on the same three product-line software benchmarks previously used for FINCH; however, {\it without} using their feature-constraint specifications.  The research question we investigate in these experiments is the following.

\emph{RQ1.G: How effective is GOLDFINCH in producing accurate results for known feature interactions?}

\subsubsection{Evaluation of GOLDFINCH on a configurable system.}

Since some feature dependencies can cause variability bugs,
we investigated whether there exists any bug or warning related to the feature dependencies that GOLDFINCH finds.  We first checked the known variability bugs  in \cite{abal2018variability}.  A variability bug is a bug that occurs due to  enabling or disabling a combination of configuration options or features. 
We focused on those 9 recorded variability bugs that involve two or more features.  
Filtering out those that are related to compilation/build issues 
yielded 4 known variability bugs.
We applied GOLDFINCH on the sample code provided for those four bugs at \cite{VBDB}. 

Table \ref{tab:knownbugs} shows our results. For three out of the four known feature-interaction configuration bugs in earlier BusyBox versions, GOLDFINCH detected a Store-Load type dependency in the sample code explaining the bug.  This experiment suggests that feature dependencies found by GOLDFINCH can be useful as indicators of unwanted feature interactions.  

The known variability bug not found by GOLDFINCH does not involve a data-flow dependency; rather, it is 
an unused variable bug, which is about the absence of data-flow rather than its existence.  It was later fixed in BusyBox by moving the declaration of 
the variable within the scope of {\tt FEATURE\_MDEV\_RENAME\_REGEXP}.

\begin{table}[ht!]
    \caption{Timing and Memory Results for Generating Store-Load (SL) and Store-Store (SS) Dependencies for the coreutils Subsystem of BusyBox 1.32.0. Timeout was set to 60 secs.}
    \label{tab:busyboxDF}
    \centering
    \scalebox{0.9}{\begin{tabular}{crrrr} \toprule
    {\bf Benchmark} & {\bf SL} & {\bf SS} & {\bf Time (s)} & {\bf Mem (MB)} \\ \toprule
    chgrp & 1 & 1 & 0.01 & 15.39 \\
    chmod & 11 & 2 & 0.01 & 15.34 \\
    chown & 2 & 0 & 0.01 & 15.14 \\
    comm & 6 & 5 &  0.01&  15.27\\
    cp & 2 & 0 & 330.70 & 219.76 \\
    cut & 8 & 1 &  0.01& 15.27 \\
    date & 6 & 1 &  0.01& 15.14 \\
    df & 7 & 1 &  0.01& 15.36 \\
    dos2unix & 1 & 0 &  65.82 & 233.69 \\
    du & 25 & 16 &  0.01& 217.25 \\
    env &  1 & 0 &  0.01& 15.34 \\
    expand & 14 & 9 &  65.23 & 218.26 \\
    expr & 13 &  4&  0.01&  15.38\\
    fold &  2& 0 &  0.01& 15.27 \\
    head &  2&  0& 0.01 &  15.33\\
    id &  29&  8 & 0.01 & 15.14 \\
    install & 0 & 0 &  324.67 & 223.35 \\
    ln &  0 & 0 &  326.94 & 219.57 \\
    ls & 8 & 10  &  0.01& 15.21 \\
    md5\_sha1\_sum   &  16& 14 & 0.01 & 15.30 \\  
    mkdir     & 0 &  0 &  64.58 & 217.73 \\
    mknod    &  1& 0 &  4.64 & 214.65 \\
    mktemp    &  1& 1 &  0.01& 15.27 \\
    nice    &  7& 1 & 0.01 & 15.28 \\ 
    nl     &  12 & 9 & 0.01 & 15.31 \\
    nproc    & 0 & 0 &  0.01& 15.30 \\
    od     &  11& 10 &  0.01& 15.36 \\
    paste   &  4 & 1 &  0.01&  15.20\\ 
    printf    & 7 &  15 &  0.01&  15.24 \\
    rm     &  9& 1 &  0.01& 15.14 \\
    shred     & 2 & 0 &  0.01& 15.20 \\
    shuf     &  1& 0 &  0.01& 15.20 \\
    sleep     &  7& 1 &  9.27 & 214.86 \\
    split    &  3& 0 &  0.01& 15.14 \\
    stat    &  1& 0 &  0.01& 15.32 \\
    stty     &  1& 0 &  0.01& 15.20 \\
    tail     &  18& 2 &  0.01& 15.20 \\
    tee   &  7& 3 &  0.01& 15.36 \\
    timeout     &  9& 1 & 0.01 & 15.28 \\
    touch    &  4& 0 & 0.01 & 15.36 \\
    tr    &  6& 1 &  0.01& 15.14 \\
    tty   &  7& 1 &  6.87 &  214.90\\
    uname     &  3&  0&  6.88 & 214.75 \\
    usleep    &  7& 1 &  0.01& 15.27 \\
    uudecode    &  11& 3 &  0.01& 15.36 \\
    wc     &  4& 11 &  64.92 & 216.60 \\
    who    &  1&  0&  6.97 & 214.85 \\
    yes &  3& 0 & 0 0.01&  15.34\\
    \bottomrule
\end{tabular} }
\end{table}    

\begin{table*}[t]
\centering
\caption{Comparison of Feature Dependency Results for PROMPT and SVF on Core Utilities of BusyBox 1.32.}
\begin{tabular}{crrrr}
\toprule
{\bf Analysis Tool} & {\bf Store-Load} & {\bf Store-Store} & {\bf Total Time (s)} & {\bf Max Memory (MB)} \\ \hline
PROMPT & 301   & 134   & 1277.85 & 233.69\\
SVF    & 402   & 960   & 3603.50 & 10844.05\\
\bottomrule
\end{tabular}%
\label{tab:prompt_svf_busybox32}%
\end{table*}%

\begin{table*}[th!]
  \caption{GOLDFINCH: Number of Feature Dependencies Across Two Types of Memory Dependencies in  Core utilities of BusyBox 1.32. }
    \centering
    \begin{footnotesize}
    \begin{tabular}{cccccc}
    \toprule
    {\bf } & {\bf } &{\bf Store-Load}& {\bf Store-Load}& {\bf Store-Store}& {\bf Store-Store}\\ 
    
    {\bf Source} & {\bf Destination} &{\bf PROMPT}& {\bf SVF}& {\bf PROMPT}& {\bf SVF}\\ \toprule
    {\bf Feature Relevant} & {\bf Not Feature Relevant} &9  & 15  & 6  & 10  \\
    {\bf Not Feature Relevant} & {\bf Feature Relevant} &8  & 3 & 2 & 4 \\
    {\bf Not Feature Relevant} & {\bf Not Feature Relevant} & 281  & 330  & 122 & 927 \\
    {\bf Feature Relevant} & {\bf Feature Relevant} & 3  & 54  & 4 & 19 \\ \hline
    \multicolumn{2}{c}{\bf Total} & 301  & 402  & 134 & 960 \\
    \bottomrule
    \end{tabular}
    \end{footnotesize}
  \label{tab:busybox_feature_dependency_stats}%
\end{table*}%

Next we evaluated our method on its planned target: a large, configurable system.  We used the latest version of BusyBox, $BusyBox 1.32.0$. We selected its coreutils both because coreutils implements BusyBox's core functionalities and because coreutils had the highest reported variability bugs within BusyBox in a recent study \cite{abal2018variability}. 
 
Table \ref{tab:busyboxDF} reports our experimental results for GOLDFINCH's detection of data-flow dependencies. The table presents the number of Store-Load and Store-Store pairs, together with timing and memory information.  We discuss the encouraging implications of the results for the scalability of our approach at the end of Section \ref{sec:results}. 

We also evaluated our approach in comparison with static analysis. To further assess our use of the dynamic symbolic-execution PROMPT tool for dependency analysis in GOLDFINCH, we compared the results from using PROMPT to those obtained by using a  static analyzer, SVF \cite{sui2016svf}.  

SVF is a state-of-the-art static analysis tool that provides a variety of pointer analysis for LLVM-based languages \cite{sui2014detecting,sui2016svf}. 
Specifically, we configured SVF to use context-sensitive pointer analysis. 

We then traversed SVF's internal representations of the Sparse Value-Flow Graph (SVFG) and the Program Assignment Graph (PAG) to find Store-Load and Store-Store dependencies, respectively. 
Tables \ref{tab:prompt_svf_busybox32}-\ref{tab:busybox_feature_dependency_stats} thus also compare data-flow dependency results for PROMPT and SVF.

We used the SVFG to capture flow-sensitive Store-Load dependencies, and had to use the Program Assignment Graph (PAG) to find out store instructions that access the same memory locations. 

While we used PROMPT to generate a Store-Store dependency for each pair of {\em consecutive} pair of store instructions that access the same memory location, 
we used the PAG created by SVF to generate a Store-Store dependency 
for any pair of store instructions that access the same memory. 
Thus, the 
set of Store-Store dependencies generated using SVF is in principle a superset of the set of Store-Store dependency  generated using PROMPT and is less precise. This is because, unlike the Store-Load dependencies, SVF does not generate Store-Store dependencies by default.
This explains why the number of Store-Store dependencies found by SVF (960) 
is much higher than those found by PROMPT (134).  
Similarly, SVF finds  
more Store-Load dependencies (402) than PROMPT (301). However, 
this is because we  configure PROMPT to abstract away the library functions and to use a symbolized return value for those with a return value \cite{yavuz2020analyzing} as a way to deal with the path explosion problem. Since library function bodies do not involve feature dependent code, SVF's reporting of Store-Load dependencies within the 
library functions is something that we can avoid using PROMPT.
As the timing results in Table \ref{tab:prompt_svf_busybox32} shows, 
GOLDFINCH scales symbolic execution to extract feature relevant models from large-scale configurable systems.

\begin{figure}[th!]
\centering
\begin{footnotesize}
 \begin{verbatim}
 store i32 %46, i32* @option_mask32, align 4, !dbg !207758, !tbaa !70234
  br label %47, !dbg !207759

; <label>:47                                      ; preds = %41, %44
  %48 = and i32 %6, 32, !dbg !207760
  %49 = icmp ne i32 %48, 0, !dbg !207760
  %50 = and i32 %6, 655360, !dbg !207762
  %51 = icmp eq i32 %50, 0, !dbg !207762
  %or.cond = or i1 %49, %51, !dbg !207767
  %.pre = load i32, i32* @option_mask32, align 4, !dbg !207768, !tbaa !70234
  br i1 %or.cond, label %._crit_edge, label %52, !dbg !207767

; <label>:52                                      ; preds = %47
  %53 = or i32 %.pre, 262144, !dbg !207770
  store i32 %53, i32* @option_mask32, align 4, !dbg !207770, !tbaa !70234
  br label %._crit_edge, !dbg !207772

._crit_edge:                                      ; preds = %47, %52
  %54 = phi i32 [ %53, %52 ], [ %.pre, %47 ], !dbg !207768
\end{verbatim}
\end{footnotesize}
\caption{An LLVM Bitcode Excerpt for the Source Code Example from coreutils/ls.c as Given in Figure \ref{fig:ssexample}.}
\label{fig:sourcelocchallenge}
\end{figure}

Both PROMPT and SVF find a true feature-relevant Store-Store dependency within coreutil's
{\tt ls.c} in BusyBox 1.32.0,  corresponding to the dependency between lines 1173-1181 shown in Figure \ref{fig:ssexample}. 
This dependency is between {\tt ENABLE\_FEATURE\_LS\_RECURSIVE} and \\
{\tt ENABLE\_FEATURE\_LS\_TIMESTAMPS \&\& ENABLE\_FEATURE\_LS\_SORTFILES}, 
as both program locations update the variable {\tt option\_mask32}. 
Although we do not think that this dependency implies a variability 
bug in that particular version of BusyBox, future changes may 
create issues. 
As an example, if a bug were to make the expressions {\tt OPT\_R} and {\tt OPT\_t} evaluate to the same value, then the second store instruction at line 1181 would set the bit that gets cleared by the first store instruction at line 1173.

PROMPT reports an additional dependency that turns out to be a false positive, involving accesses to different fields of the 
same {\tt struct} type object. The cause of this false positive is that  the KLEE symbolic 
execution engine used in PROMPT represents the entire {\tt struct} object as a 
single memory object. In future work, we will incorporate offset 
information into our analysis to eliminate such false positives.

It is remarkable that both PROMPT and SVF missed the Store-Load 
dependency within coreutil's
{\tt ls.c} in BusyBox 1.32.0,  corresponding to the dependency between lines 1173-1181 in Figure \ref{fig:ssexample}. 
A deeper investigation into why both tools missed this dependency revealed 
precision issues related to source line number generation during compilation.

Figure \ref{fig:sourcelocchallenge} demonstrates one source of imprecision in GOLDFINCH due to the optimizations performed by the compiler, 
specifically, the load instructions related to the {\tt option\_mask32} 
variable on lines 1181 and 1187 in Figure \ref{fig:ssexample}.
These two load instructions are used in a {\tt phi} instruction so that, 
depending on the evaluation of the branch instruction on line 1180 
in Figure \ref{fig:ssexample},
either the result of the instruction at line 10 or line 14 in Figure 
\ref{fig:sourcelocchallenge} gets used. The issue is that, 
the source line reported for the instruction at line 10 is the same 
as that reported for the instruction at line 17, which happens to be 1187 and does not get 
included within the scope of any feature. However, the source line 
for the instruction at line 10 is actually 1181. 
This causes GOLDFINCH to 
miss this feature-relevant dependency as the destination tuple 
seems not be related to any feature.  
It illustrates why source line numbers may not always be precisely reflected to the results of GOLDFINCH, which performs dependency analysis  at the intermediate-representation level.

\begin{table}[th!]
    \caption{Timing and Memory Results for Generating Store-Load (SL) and Store-Store (SS) Dependencies for the Product Lines. Timeout was set to 60 secs.}
    \label{tab:prodDF}
    \centering
    \begin{footnotesize}
    \begin{tabular}{crrrr} \toprule
    {\bf Benchmark} & {\bf SL} & {\bf SS} & {\bf Time (s)} & {\bf Mem (MB)} \\ \toprule
        elevator\_spec1 & 475 & 426 & 100.00 & 66.49\\
        elevator\_spec2 & 352 & 178 & 100.00 & 46.78\\
        elevator\_spec3 & 437 & 507 & 100.00 & 87.04 \\
        elevator\_spec9 & 356 & 238 & 100.00 & 42.15\\
        elevator\_spec13 & 350 & 146 & 100.00 & 36.34\\
        elevator\_spec14 & 444 & 445 & 100.00 & 50.02\\
        email\_spec0 & 275 & 17 & 100.00 & 237.07\\
        email\_spec1 & 265 & 17 & 100.00 & 250.10 \\
        email\_spec3 & 287 & 25 & 100.00 & 226.56\\
        email\_spec4 & 266 & 28 & 100.00 & 208.84\\
        email\_spec6 & 258 & 17 & 100.00 & 215.13\\
        email\_spec7 & 254 & 16 & 100.00 & 175.22\\
        email\_spec8 & 260 & 17 & 100.00 & 234.68\\
        email\_spec9 & 253 & 18 & 100.00 & 215.09 \\
        email\_spec11 & 293 & 19 & 100.00 & 216.24\\
        email\_spec27 & 289 & 23 & 100.00 & 239.80 \\
        minepump\_spec1 & 56 & 9 & 100.00 & 389.52\\
        minepump\_spec2 & 58 & 16 & 100.00 & 379.85\\
        minepump\_spec3 & 65 & 9 & 100.00 & 385.63\\
        minepump\_spec4 & 55 & 9 & 100.00 & 373.12\\
        minepump\_spec5 & 60 & 12 & 100.00 & 353.97\\ 
        \bottomrule
    \end{tabular}
    \end{footnotesize}
\end{table}

\begin{table*}[th!]
  \caption{GOLDFINCH: Result of Data Flow Dependencies in Finding Unwanted Feature Interaction without Specifications in Three Small Software Product Lines}
    \centering
    \begin{footnotesize}
    \begin{tabular}{ccccc} \toprule
    {\bf SPL name} &{\bf \# Data Flow }& {\bf \# Unwanted Feature }& {\bf\# Detected Unwanted Feature }& {\bf Time(S)}\\
    
    {} &{\bf Dependency}& {\bf Interactions}& {\bf  Interactions}& {}\\
    
    \toprule
    {\bf Email} & 2897 & 10  & 10 & 33.52 \\
    {\bf Elevator} & 4354 & 5 & 4 & 12.21 \\
    {\bf Mine Pump} & 349  & 4  & 3 & 3.03 \\
    \bottomrule
    \end{tabular}
    \end{footnotesize}
  \label{tab:productline_feature_dependency}%
\end{table*}%

Although the number of interactions found is small, our results are consistent with prior findings. For example, Garvin, Cohen and Dwyer showed that only a small set of feature combinations in a system's configuration space are associated with failures \cite{garvin2013failure}.  The relative rarity of unwanted feature interactions may be part of the reason why constraints on combinations of features often remain unspecified or are not updated as options evolve \cite{Cashman18}.   
Here, manual inspection of the dependencies found by GOLDFINCH indicate that either a distinct 
set of memory locations are accessed in each feature or 
accesses to the same memory location at different program locations 
correspond to the same feature combination.

\subsubsection{Evaluation of GOLDFINCH on software product lines}

We also evaluated our feature dependency method GOLDFINCH  on the same three small software product lines--Email, Elevator and Mine Pump--that we used for our evaluation of FINCH. This enabled us to compare GOLDFINCH's  results {\emph without} access to feature-constraint specifications with FINCH's results {\emph with} access to feature-constraint specifications. To make this comparison we thus used the known unwanted feature interactions for the software product lines \cite{apel2013feature} only as our ground truth.

As with our evaluation of GOLDFINCH on BusyBox, we used the framework described in Fig.  \ref{fig:framework_withou_specs}.   
We again use data flow interactions to discover feature dependencies, that is, the features that are dependent on the same location of memory.  As shown in Step 1 of Fig. \ref{fig:framework_withou_specs}, the PROMPT tool extracts the feature-relevant model from the product lines' source code. 
Table \ref{tab:prodDF} presents the number of Store-Load and Store-Store 
pairs and the timing information.

As shown in Step 2 of Fig. \ref{fig:framework_withou_specs}, we automatically locate the feature-relevant data flow by analyzing the feature dependencies of the two memory dependency types, Store-Store and Store-Load for each of the three product lines.  Since the implementation of features in these product lines is not based on the \#ifdef C preprocessor, we do not use the cppcheck tool\cite{hunsen2016preprocessor} to locate features, as we had in BusyBox. Instead, we use the names of the functions where the data flow interactions occur to automatically locate the  features implemented by those functions.

For example, the following two functions in the Email product line access the same memory region, with ${printMail\_\_role\__Sign\_source\_sl\_s}$ storing to the same variable that \\ ${printMail\_\_role\_\_Verify\_dest\_sl\_l}$ loads from it. The name of the function identifies the name of its related feature, i.e.,  ${printMail\_\_role\_\_Sign()}$ belongs to the Sign feature and \\ ${printMail\_\_role\_\_verify()}$ belongs to the Verify feature. The type of memory dependency here is Store\_Load (sl), with the Sign feature being the Source and store (s) the Value; the Verify feature being the Destination and load (l) being from the variable that the Sign feature stored to it. 

Step 3 of GOLDFINCH, as shown in Fig. \ref{fig:framework_withou_specs}, uses association rule mining, namely the Apriori algorithm \cite{borgelt2012frequent}, to learn the frequent item set of features that interact together in the feature dependency data.   The feature dependencies extracted by PROMPT, once encoded, serve as the input to the Apriori unsupervised learning in order to find the most relevant features.
Here, GOLDFINCH correctly detects that the Sign and Verify features have a potential feature interaction.  This is a true positive since it is a known unwanted feature interaction, as reported in Table \ref{tab:known_fi_email}.

As shown in Step 4 of Fig. \ref{fig:framework_withou_specs}, GOLDFINCH provides a set of association rules detected by the Apriori unsupervised learning model to developers to inform them of possible feature interaction pairs of concern. For our Sign/Verify example the following rule is output: \\
$\{printMail\_\_role\__Sign\_source\_sl\_s\} 
\iff 
\{printMail\_\_role\_\_Verify\_dest\_sl\_l\}$.

Table \ref{tab:productline_feature_dependency} summarizes the results from our experimental evaluation of GOLDFINCH for each of the three software product lines, Email, Elevator and Mine Pump.  The second column of the table shows the number of known unwanted feature interactions for each product line \cite{apel2013feature}. These specifications are used as the ground truth for our evaluation of the results. 
The third column shows the number of these known feature interactions that GOLDFINCH finds.  For the Email software product line, GOLDFINCH finds all ten of the known unwanted feature interactions (100\%) described in \cite{apel2013feature}. For the Elevator product line, GOLDFINCH finds four out of the five (80\%)  unwanted feature interactions in it. For the Mine Pump product line, GOLDFINCH finds three out of its four (75\%) unwanted feature interactions.
As we can see in Table 12, some constraint specifications are not implemented in the code. As a consequence, we could not detect two unwanted feature interactions in Elevator and Mine Pump.

The results in Section \ref{sec:eval_with} clearly show the advantage of using FINCH's supervised learning when unwanted feature interactions {\it are specified and available}. Moreover, the results in Section \ref{sec:eval_without} show that, when unwanted feature {\it specifications are not available},  GOLDFINCH's unsupervised learning successfully discovered many feature dependencies of concern that otherwise could elude developers. 
For the many highly configurable, real-world systems without access to feature-constraint specifications, GOLDFINCH offers an effective way to surface and increase understanding of unrecognized feature dependencies.

{\bf Finding: When feature-constraint specifications were not available, our method efficiently discovered
the same feature relevant data dependency that static analysis approach discovered, as well as

17 of the 19 unwanted feature interactions in  three product lines.}

\emph{RQ3.G:  How scalable is  the  performance  of  our  approach when specifications are unavailable?}
 We evaluate the scalability of GOLDFINCH based on  its handling of a large number of features in software product lines and of options in a highly configurable system. 
As shown in Table \ref{tab:productline_feature_dependency}, 
GOLDFINCH detects unwanted feature interactions in three small software product lines of six to ten features in 3 to 34 seconds.  As shown in Table \ref{tab:prompt_svf_busybox32}, 
GOLDFINCH performs comparably with SVF, a static analysis tool, 
in analyzing the core utilities of a real-world configurable system 

with 139 features. 
Since SVF is a whole-program analysis tool, the time we report in 
Table \ref{tab:prompt_svf_busybox32} is the total time SVF spends 
on the BusyBox code base, which is approximately 3 times the total 
time spent by PROMPT on the coreutils subsystem of BusyBox. 
Similarly, the  memory presented in  Table \ref{tab:prompt_svf_busybox32} for
SVF is for the analysis of the entire BusyBox code base, which is 
50 times the memory used by PROMPT for the analysis of the coreutils subsystem. 

This is significant in that, although SVF's static analysis has better coverage than PROMPT's 
API modeling based analysis, PROMPT detects the same 
feature relevant dependency in the coreutils subsystem of BusyBox 1.32.0 as found by SVF while achieving a performance that is comparable in terms of running time and more optimized in terms of memory usage.

{\bf Finding: Results from our experiments indicate that our GOLDFINCH approach, for use when feature-constraint specifications are not available, is scalable in terms of the number of features and code size it can handle.}

\section{Related Work}
\label{sec:related}

Previous research on using machine learning to detect feature interactions has aimed primarily at coverage testing of configurable systems \cite{nguyen2016igen}, at finding faulty configurations without incurring the cost of testing \cite{temple2016using}, or on the effect of feature interactions on performance \cite{bacciu2015using, kolesnikov2019relation,velez2019configcrusher}.  Recent work has suggested that combining information from learning-aided configuration with information from experts may further improve results \cite{amand2019towards}.

A study of feature interactions in the product-line literature by Soares, et al. \cite{soares2018feature} found that 43\% of the papers aimed to understand feature interaction at early stages of the software life cycle.  Among this 43\%, the majority used formal methods, especially feature-aware verification to automate detection of interactions \cite{apel2011detection, apel2013feature}, and model checking to measure behavioral changes when a new feature is added \cite{atlee2015measuring}, as well as to detect conflicts among features \cite{calder2006feature, beidu2019detecting}. However, such formal models are costly to create and typically not available for most real-world systems.

Our approach to dealing with the feature interaction problem is the first we are aware of that uses symbolic-execution-guided machine learning on a model built using data from prior products in a software product line to detect unwanted feature interactions in a new product. Similar to other approaches that use call sequence information \cite{raychev2014code,yessenov2017demomatch},  
our approach also learns some type of semantic association among the functions. However,
unlike in these works, we consider systems in which samples of correct usage of the functions and 
features may not be available. Our use of symbolic constraints in learning differs from the approach in \cite{fowze2019proxray} as our constraints are not on the input variables and, instead, involve internal state of the computation, which is important for detecting unwanted feature interactions.

Regarding feature interactions in highly configurable systems without specifications, Soares, et al. \cite{soares2018exploring} used dynamic analysis based on the data and control flow execution of Java-based configurable systems. Our work reported here differs in that we focus on analyzing data flow and static analysis of C-based configurable systems and do not require availability of test cases.

Kolesnikov, et al. \cite{kolesnikov2019relation} investigated  the relation of control flow and performance feature interactions in predicting the performance of a highly configurable system in terms of timing. Our work differs since we use data flow interaction to predict feature interactions in a highly configurable system. 
Rhein, et al. \cite{rhein2018variability} proposed a variability aware static analysis technique based on seven control and data flow analyses. They reported that their technique out-performed  sample-based, static-analysis approaches in  terms of execution times and  potential bugs found. Our study differs in its use of highly accessible and  general-purpose static analysis tools including PROMPT \cite{yavuz2020tutorial} and SVF \cite{sui2014detecting,sui2016svf} to capture the data-flow interactions in highly configurable systems. 

Our attention to feature interactions is also shared by combinatorial interaction testing  (CIT) techniques \cite{yilmaz2014moving}, which have been applied to the testing of both  software product lines and highly configurable systems. CIT identifies a subset of features to be considered in combination to achieve, most commonly, pairwise coverage \cite{lopez2015first}.  It is a form of sampling, informed by constraints on allowable feature combinations, that reduces the number of tests needed.  Our approach differs from CIT in our use of symbolic-execution-guided learning and in  identifying interactions whether or not feature-constraint specifications exist.

\section{Discussion}
\label{sec:conclusion}

In this section we describe use cases for FINCH and its generalization in GOLDFINCH,  
and discuss threats to validity.

{\it Use Cases.} We have proposed a method to automatically target feature  interactions, whether or not unwanted  feature interactions are explicitly specified.  The envisioned user of our method  is a developer implementing a new product in a software product line or  new or changed options in a highly configurable system.   Feature interactions have been shown to be hard for developers to detect in a new product \cite{kruger2019effects}.  

Even where constraints on feature combinations have been documented, that information may not be accessible to developers, may be obsolete, or may not match the code.   Moreover, the rationales for avoiding certain feature interactions may be undocumented or subtle, requiring additional domain knowledge beyond the developer's expertise, and thus be not taken into account in the source code.   The features involved in an unwanted interaction often are in components that are distant from each other and may be the responsibility of different individual developers or teams, further complicating their discovery and analysis. 

Moreover, atypical or rare combinations of features may not have been thoroughly analyzed or tested, especially if they were added ad-hoc, for example, in response to an urgent customer need. 

There are thus three use cases that motivate our work.  The first use case is to answer a developer's query 
as to whether a change enables features to interact so as to violate a specified feature constraint. In this case, the developer uses the output from FINCH to check that the planned pairing of 
features in a new product still satisfies the relevant specification.  This helps validate the introduction of a new feature combination earlier in development.

The second use case is to give developers information that helps them test for unwanted feature interactions. The output from our learning algorithms identifies combinations of features and paths to prioritize in probing for unwanted feature interactions. Our method also gives testers the feature-dependency information they need regarding which configurations to cover in tests.

The third use case is to improve program comprehension. Feature interactions often cause problems for program comprehension \cite{kruger2019effects}.  Improved automatic detection of feature interactions improves program understanding toward reducing bugs and speeding bug repair of unwanted interaction behavior. 
This includes information both about unwanted feature interactions known to have occurred in prior products and the learning-based detection of new feature dependencies and interactions that our automated method provides to the developers. 
An additional potential use of our method is to reduce software aging. Feature-rich software systems such as software product lines and highly configurable software are subject to aging over time \cite {parnas1994software, cotroneo2014survey},  with performance degradations and increased failure rates.  A known cause of software aging in these systems is the introduction of new features that interact in unforeseen ways with existing features \cite{johnsson2000quantifying, siegmund2012spl, garvin2013failure}.  A strength of GOLDFINCH is that, for the many feature-rich systems lacking feature-constraint specifications, it can help reduce this cause of aging by pinpointing problematic feature dependencies meriting developers’ further attention.

{\it Threats to Validity.}  An external threat to validity is that we used only three small product lines from the literature and a large, real-world subsystem to evaluate our approach. However, these software systems are all well-studied and open-sourced, are in different domains, and have a variety of features and unwanted feature interactions.  Our results showed similarities in the models' accuracy and performance, and indicated the feasibility of our approach.
An internal threat to validity is that the specifications for the product lines might be incorrect, leading to incorrect results regarding which interactions are unwanted. 
However, the product lines we used have been used by other researchers, and we reviewed those papers as well as the code to ensure that the feature interactions considered here to be faulty in fact were problematic.  A few papers have added other features and feature interactions to the email product line beyond those that appear in this paper; however, we restricted ourselves to those interactions that are more standard across the literature.  Another internal threat to validity is that we only consider feature interaction pairs. However, as previously noted, pairwise interactions have been found to be the most common, and it has been shown that it is exceedingly rare to find a 3-way feature interaction that is not detectable as a 2-way interaction \cite{calder2006feature}.

\section{Conclusion}
\label{sec:summary}
In this section we summarize our results and offer concluding remarks regarding the contributions of the paper.   

Our proposed approach is shown in the experiments described here to be effective in producing accurate results for unwanted feature interactions both with and without the availability of feature-constraint specifications.  When feature-constraint specifications are available, results from FINCH show that we are able to classify the data from both stack traces and path constraints with high balanced accuracy to predict unwanted feature interactions in three small software product lines. Prediction is very fast, and our approach predicts the actual failure paths related to unwanted feature interactions in 0.007-0.6 seconds. The models built using SVM and combined data (both stack traces and path constraints) have high balanced accuracy across all three product lines.

 Our evaluation of FINCH shows that it accurately predicts some new feature interactions, based on its learned model of existing feature interactions. Moreover, it detects failure paths related to unwanted feature interactions with only partial data, using only 75\% of functions called and atomic path constraints in three product-line case studies. The finding that we can achieve the same high accuracy with partial data is promising both for faster prediction of unwanted feature interactions in  product lines and for  steering the symbolic execution engine towards the failure paths.
 
In real-world applications there are often no feature-constraint specifications available  \cite{soares2018exploring}. 

For these cases, we extend our approach to seek unwanted feature interactions through automated detection of feature-relevant data-flow dependencies. 

Our generalized approach, GOLDFINCH, is able to detect a Store-Store type dependency between two features in BusyBox's coreutilies subsystem that previously were involved in an unwanted feature interaction.

While results confirm the advantage of using specifications when they are available, GOLDFINCH detects 17 of the 19 unwanted feature interactions across three software product lines even without access to feature-constraint specifications.

 Results reported here indicate that  GOLDFINCH is scalable in terms of the number of features and code size it can handle.  In experiments comparing our automated method with SVF, a static analysis tool, our method has comparable running-time performance and more optimized memory usage.

In summary,  the automated learning-based method we describe here has the potential to provide developers with earlier and better insights into unwanted feature interactions in feature-rich systems such as product lines and highly configurable systems.  On projects where specifications of constraints on feature combinations already exist,

we use symbolic execution-guided supervised learning to automatically predict unwanted feature interactions in a new or evolving product.  On projects where such specifications are not available, 

we use dynamic symbolic-execution-guided program analysis and association rule mining to automatically infer potentially problematic feature dependencies meriting developers' attention.

\section{Acknowledgements}
This work was supported in part by the National Science Foundation under grants CCF-1513717 and CNS-1942235.


\bibliographystyle{plain}
\bibliography{reference}


\end{document}